\documentclass[preprint,showpacs,preprintnumbers,amsmath,amssymb]{revtex4}
\usepackage{graphicx}
\usepackage{dcolumn}
\usepackage{bm}
\usepackage{color}
\usepackage{hhline}

\begin{document}

\title{Simulation of spin-polarized scanning tunneling spectroscopy on complex magnetic surfaces:
       Case of a Cr monolayer on Ag(111)}

\author{Kriszti\'an Palot\'as}
\email{palotas@phy.bme.hu}
\affiliation{Budapest University of Technology and Economics,
Department of Theoretical Physics, Budafoki \'ut 8., H-1111 Budapest, Hungary}

\author{Werner A. Hofer}
\affiliation{University of Liverpool, Surface Science Research Centre, L69 3BX Liverpool, United Kingdom}

\author{L\'aszl\'o Szunyogh}
\affiliation{Budapest University of Technology and Economics,
Department of Theoretical Physics and Condensed Matter Research Group of the Hungarian Academy of Sciences,
Budafoki \'ut 8., H-1111 Budapest, Hungary}

\date{\today}

\begin{abstract}

We propose a computationally efficient atom-superposition-based method for simulating spin-polarized scanning tunneling
spectroscopy (SP-STS) on complex magnetic surfaces based on the sample and tip electronic structures obtained from first
principles. We go beyond the commonly used local density of states (LDOS) approximation for the differential conductance, dI/dV.
The capabilities of our approach are illustrated for a Cr monolayer on a Ag(111) surface in a noncollinear magnetic state.
We find evidence that the simulated tunneling spectra and magnetic asymmetries are sensitive to the tip electronic structure,
and we analyze the contributing terms. Related to SP-STS experiments, we show a way to simulate two-dimensional
differential conductance maps and qualitatively correct effective spin polarization maps on a constant current contour above a
magnetic surface.

\end{abstract}


\maketitle

\section{Introduction}

The scanning tunneling microscope (STM) and its spectroscopic mode (STS) proved to be extremely useful for studying local physical
and chemical phenomena on surfaces since the invention of the STM 30 years ago \cite{binnig1,binnig2}.
The progress of experimental techniques in the last two decades was remarkable, thus, more sophisticated theoretical models and
simulation tools are needed to explain all relevant details of electron tunneling transport measurements
\cite{hofer03rmp,hofer03pssci}. STS theory and applications are recently focused on extracting surface local electronic properties
from experimental differential conductance ($dI/dV$) data \cite{ukraintsev96,koslowski07,passoni09,ziegler09,koslowski09}.
The role of the tip electronic structure has been identified to be crucial on the $dI/dV$ tunneling spectra, see
e.g.\ Refs.\ \cite{passoni09,kwapinski10}, and a theoretical method has been proposed to separate the tip and sample
contributions to STS \cite{hofer05sts}.

An emerging research field in surface science is the investigation of magnetism at the nanoscale and atomic scale with the aim of
achieving ultrahigh information density for data storage purposes \cite{ultrahigh,weiss05}. Spin-polarized scanning
tunneling microscopy (SP-STM) \cite{bode03review} is admittedly an important tool for studying magnetism on surfaces.
Recent experimental advances using this technique allow the investigation of complex magnetic structures (frustrated
antiferromagnets, spin spirals, skyrmion lattices, etc.) \cite{wiesendanger09review,wulfhekel10review,serrate10,heinze11skyrmion}.
Spin-polarized scanning tunneling spectroscopy (SP-STS) has recently been used to find inversion of spin polarization above
magnetic adatoms \cite{yayon07,heinrich09,zhou10}, and the effect has been explained theoretically \cite{ferriani10tip}.
Furthermore, SP-STS is useful to study atomic magnetism \cite{wiebe11}, many-body effects on substrate-supported
adatoms \cite{neel10kondo}, or magnetic interactions between adatoms \cite{ternes09sts} as well.
The effect of differently magnetized surface regions on SP-STS has also been reported
\cite{schouteden08,heinrich10}, and the role of tip effects on SP-STS \cite{rodary09,palotas11sts} and on achieving giant magnetic
contrast \cite{hofer08tipH} have also been highlighted.

Our work is concerned with the presentation of an efficient simulation method for SP-STS based on first principles
electronic structure data. We extend our atom-superposition-based method \cite{palotas11sts,palotas11stm} in
the spin-polarized Tersoff-Hamann framework \cite{wortmann01} for simulating SP-STS by including the bias dependent
background and tip-derivative terms into the calculated differential conductance following Passoni et al.\ \cite{passoni09}.
The method is computationally cheap and it can be applied using results of any ab initio electronic structure code.
The main advance of our tunneling model is the inclusion of the tip electronic structure, which is neglected in
Refs.\ \cite{wortmann01,heinze06}, and it is only taken into account in a model way in Ref.\ \cite{passoni09}. Our method,
based on first principles calculation of the tip electronic structure, enables to study tip effects on the SP-STS spectra.
Taking a prototype frustrated hexagonal antiferromagnetic system, a Cr monolayer on Ag(111) in a noncollinear
magnetic $120^{\circ}$ N\'eel state, we simulate differential conductance tunneling spectra and magnetic asymmetries to illustrate
the applicability of our method, and we analyze the contributing terms.
Note that a three-dimensional (3D) approach to STS has been presented recently, that is applicable to nonmagnetic
systems only, and it takes into account the symmetry of the tip states but not the electronic structure of the tip apex
\cite{donati11}.
Our model is also a 3D approach in the sense that we sum up contributions from individual transitions between the tip apex atom
and each of the surface atoms assuming the one-dimensional (1D) Wentzel-Kramers-Brillouin (WKB) approximation for electron
tunneling processes in all these transitions, thus we call it a 3D WKB approach.

The paper is organized as follows: The atom-superposition theoretical model of SP-STS is presented in section
\ref{sec_spsts}. As an application, we investigate the surface of one monolayer (ML) Cr on Ag(111) in section \ref{sec_res}.
We simulate differential conductance tunneling spectra and magnetic asymmetries with two tip models, and we analyze the
contributing terms to $dI/dV$.
Moreover, we show simulation results of bias dependent two-dimensional (2D) differential conductance and
qualitatively correct effective spin polarization maps following a constant current contour above the surface,
corresponding to a standard experimental setup. Our conclusions are found in section \ref{sec_conc}. Finally,
in appendix \ref{sec_app}, we report the 1D WKB theory of STS, and give alternative expressions for the $dI/dV$.

\section{Theoretical model of atom-superposition SP-STS}
\label{sec_spsts}

The 1D WKB theory for nonmagnetic STS is a well established approach \cite{ukraintsev96,passoni07}, see appendix \ref{sec_app}.
Here, we extend it to spin-polarized systems, and adapt it to an atom superposition framework, which enables a computationally
inexpensive calculation of tunneling properties based on first principles electronic structure data.

In magnetic STM junctions, the total tunneling current can be decomposed into non-spinpolarized (TOPO) and
spin-polarized (MAGN) parts \cite{wortmann01,yang02,smith04,heinze06},
\begin{equation}
I_{TOTAL}=I_{TOPO}+I_{MAGN}.
\end{equation}
Following the spin-polarized Tersoff-Hamann model \cite{wortmann01} and its adaptation to the atom superposition framework
\cite{heinze06,palotas11stm}, the magnetic contribution to the simple expression of the differential conductance
at a given energy is proportional to the scalar product of the tip and sample magnetic density of states (DOS)
vectors, $\underline{m}_T(E)$ and $\underline{m}_S(E)$, respectively,
\begin{equation}
\frac{dI_{MAGN}}{dU}(E)\propto \underline{m}_T(E)\underline{m}_S(E).
\end{equation}
Thus, the spin-polarized parts of $dI/dV$ can similarly be calculated within the 1D WKB approximation as reported
in appendix \ref{sec_app}, just replacing $n_T(E)n_S(E)$ by $\underline{m}_T(E)\underline{m}_S(E)$.

We formulate the tunneling current, the differential conductance and their TOPO and MAGN parts
within the atom superposition framework following Ref.\ \cite{palotas11stm}. Here, we assume that electrons tunnel through one
tip apex atom, and we sum up contributions from individual transitions between this apex atom and each of the surface atoms
assuming the 1D WKB approximation for electron tunneling processes in all these transitions.
The tunneling current at the tip position $\underline{R}_{TIP}(x,y,z)$ and at bias voltage $V$ is given by
\begin{equation}
\label{Eq_curr}
I_{TOTAL}(x,y,z,V)=I_{TOPO}(x,y,z,V)+I_{MAGN}(x,y,z,V),
\end{equation}
where the TOPO and MAGN terms are formally given as
\begin{eqnarray}
I_{TOPO}(x,y,z,V)&=&\int_{0}^{V}\frac{dI_{TOPO}}{dU}(x,y,z,U,V)dU\\
I_{MAGN}(x,y,z,V)&=&\int_{0}^{V}\frac{dI_{MAGN}}{dU}(x,y,z,U,V)dU.
\end{eqnarray}
The integrands are the so-called virtual differential conductances,
\begin{eqnarray}
\label{Eq_diduTOPO}
\frac{dI_{TOPO}}{dU}(x,y,z,U,V)&=&\varepsilon^2\frac{e^2}{h}\sum_{\alpha}e^{-2\kappa(U,V)d_{\alpha}(x,y,z)}n_T(E_F^T+eU-eV)n_S^{\alpha}(E_F^S+eU)\\
\frac{dI_{MAGN}}{dU}(x,y,z,U,V)&=&\varepsilon^2\frac{e^2}{h}\sum_{\alpha}e^{-2\kappa(U,V)d_{\alpha}(x,y,z)}\underline{m}_T(E_F^T+eU-eV)\underline{m}_S^{\alpha}(E_F^S+eU).
\label{Eq_diduMAGN}
\end{eqnarray}
Here, $e$ is the elementary charge, $h$ the Planck constant, and $E_F^T$ and $E_F^S$ the Fermi energies of the tip
and the sample surface, respectively. $\varepsilon^2e^2/h$ ensures that the $dI/dU$ is correctly measured in the units of $A/V$.
$\varepsilon$ has been chosen to 1 eV, but its actual value has to be determined comparing simulation results to experiments.
The sum over $\alpha$ corresponds to the atomic superposition and has to be carried out, in principle, over all surface atoms.
Convergence tests, however, showed that including a relatively small number of atoms in the sum provides converged $dI/dU$ values
\cite{palotas11sts}. The tip and sample electronic structures are included into this model via projected DOS (PDOS) onto the atoms,
i.e.\ $n_T(E)$ and $n_S^{\alpha}(E)$ denote projected charge DOS onto the tip apex and the $\alpha$th surface atom, respectively,
while $\underline{m}_T(E)$ and $\underline{m}_S^{\alpha}(E)$ are projected magnetization DOS vectors onto the corresponding
atomic spheres. They can be obtained from collinear or noncollinear electronic structure calculations \cite{palotas11stm}.
In the present work we determine the noncollinear PDOS for the sample surface and we use a collinear PDOS for a model CrFe tip
\cite{ferriani10tip}.

The transmission probability for electrons tunneling between states of atom $\alpha$ on the surface and the tip apex is
of the simple form,
\begin{equation}
\label{Eq_Transmission}
T(E_F^S+eU,V,d_{\alpha}(x,y,z))=e^{-2\kappa(U,V)d_{\alpha}(x,y,z)}.
\end{equation}
This corresponds to a spherical exponential decay of the electron wavefunctions.
Here, $d_{\alpha}(x,y,z)=\left|\underline{R}_{TIP}(x,y,z)-\underline{R}_{\alpha}\right|$ is the distance between the tip apex and
the surface atom labeled by $\alpha$ with position vector $\underline{R}_{\alpha}$. Assuming an effective rectangular potential
barrier between the tip apex and each surface atom, the vacuum decay $\kappa$ can be written as
\begin{equation}
\label{Eq_kappa_WKB}
\kappa(U,V)=\frac{1}{\hbar}\sqrt{2m\left(\frac{\phi_S+\phi_T+eV}{2}-eU\right)},
\end{equation}
where the electron's mass is $m$, $\hbar$ is the reduced Planck constant, and $\phi_S$ and $\phi_T$ are the average electron
workfunction of the sample surface and the local electron workfunction of the tip apex, respectively. The method of determining
the electron workfunctions is reported in Ref.\ \cite{palotas11stm}. $\kappa$ is treated within the independent orbital
approximation \cite{tersoff83,tersoff85,heinze06}, which means that the same spherical decay is used for all type of orbitals.
The interpretation of our simulation results with quantitative reliability compared to experiments has to be taken
with care due to this approximation. However,
extension of our model to take into account orbital dependent vacuum decay following Chen's work \cite{chen90} is
planned in the future, which is relevant for a more advanced description of tunneling between directional orbitals.

Moreover, in our model, the electron charge and magnetization local density of states above the sample surface in the
vacuum, $n_{LDOS}$ and $\underline{m}_{LDOS}$, respectively, are approximated by the following expressions:
\begin{eqnarray}
\label{Eq_nLDOS}
n_{LDOS}(x,y,z,E_F^S+eU)=\sum_{\alpha}e^{-2\kappa(U)d_{\alpha}(x,y,z)}n_S^{\alpha}(E_F^S+eU)\\
\underline{m}_{LDOS}(x,y,z,E_F^S+eU)=\sum_{\alpha}e^{-2\kappa(U)d_{\alpha}(x,y,z)}\underline{m}_S^{\alpha}(E_F^S+eU).
\label{Eq_mLDOS}
\end{eqnarray}
with
\begin{equation}
\kappa(U)=\frac{1}{\hbar}\sqrt{2m\left(\phi_S-eU\right)}.
\label{Eq_kappa_TH}
\end{equation}
Note that the exact LDOS can be obtained by explicitly calculating the decay of the electron states into the vacuum taking their
orbital symmetry into account as well, not via such a simple 3D WKB model. Our approach, however,
has computational advantages as discussed in Ref.\ \cite{palotas11stm}.

Similarly to the tunneling current, the physical differential conductance can be decomposed into non-spinpolarized (TOPO)
and spin-polarized (MAGN) parts and it can be written at the tip position $\underline{R}_{TIP}(x,y,z)$ and at bias voltage $V$ as
\begin{equation}
\label{Eq_didv_Decomp}
\frac{dI_{TOTAL}}{dV}(x,y,z,V)=\frac{dI_{TOPO}}{dV}(x,y,z,V)+\frac{dI_{MAGN}}{dV}(x,y,z,V),
\end{equation}
where the contributions are given as [see Eq.(\ref{Eq_didv2}) in appendix \ref{sec_app}]
\begin{eqnarray}
\label{Eq_didvTOPO}
\frac{dI_{TOPO}}{dV}(x,y,z,V)&=&\frac{dI_{TOPO}}{dU}(x,y,z,V,V)+B_{TOPO}(x,y,z,V)+D_T^{TOPO}(x,y,z,V)\\
\frac{dI_{MAGN}}{dV}(x,y,z,V)&=&\frac{dI_{MAGN}}{dU}(x,y,z,V,V)+B_{MAGN}(x,y,z,V)+D_T^{MAGN}(x,y,z,V).
\label{Eq_didvMAGN}
\end{eqnarray}
Here, $B$ and $D_T$ are the background and tip-derivative terms, respectively, see appendix \ref{sec_app}.
The background term, which contains the bias-derivative of the transmission function, is usually taken into account in recent
STS theories \cite{passoni07,passoni09,donati11}, while the tip-derivative term containing the energy derivative of the tip
DOS is rarely considered in the recent literature.
Obviously, the total differential conductance can also be written in the same structure,
\begin{equation}
\label{Eq_didv_3terms}
\frac{dI_{TOTAL}}{dV}(x,y,z,V)=\frac{dI}{dU}(x,y,z,V,V)+B(x,y,z,V)+D_T(x,y,z,V)
\end{equation}
with
\begin{eqnarray}
\frac{dI}{dU}(x,y,z,V,V)&=&\frac{dI_{TOPO}}{dU}(x,y,z,V,V)+\frac{dI_{MAGN}}{dU}(x,y,z,V,V)\\
B(x,y,z,V)&=&B_{TOPO}(x,y,z,V)+B_{MAGN}(x,y,z,V)\\
D_T(x,y,z,V)&=&D_T^{TOPO}(x,y,z,V)+D_T^{MAGN}(x,y,z,V).
\end{eqnarray}

In order to calculate the background term, we need the bias-derivative of the transmission function.
Using Eq.(\ref{Eq_Transmission}) and the given form of the vacuum decay in Eq.(\ref{Eq_kappa_WKB}), we obtain
\begin{equation}
\label{Eq_TransmissionDer}
\frac{\partial T}{\partial V}(E_F^S+eU,V,d_{\alpha}(x,y,z))=-\frac{me}{\hbar^2}d_{\alpha}(x,y,z)\frac{T(E_F^S+eU,V,d_{\alpha}(x,y,z))}{\kappa(U,V)}.
\end{equation}
Considering this, and the corresponding $dI/dV$ components in the 1D WKB model, Eqs.\ (\ref{Eq_B}) and
(\ref{Eq_DT}) in appendix \ref{sec_app}, the background and the tip-derivative contributions can be written as
\begin{eqnarray}
\label{Eq_B_TOPO}
B_{TOPO}(x,y,z,V)&=&-\varepsilon^2\frac{me^3}{2\pi\hbar^3}\\
&\times&\sum_{\alpha}d_{\alpha}(x,y,z)\int_{0}^{V}\frac{e^{-2\kappa(U,V)d_{\alpha}(x,y,z)}}{\kappa(U,V)}n_T(E_F^T+eU-eV)n_S^{\alpha}(E_F^S+eU)dU\nonumber\\
\label{Eq_B_MAGN}
B_{MAGN}(x,y,z,V)&=&-\varepsilon^2\frac{me^3}{2\pi\hbar^3}\\
&\times&\sum_{\alpha}d_{\alpha}(x,y,z)\int_{0}^{V}\frac{e^{-2\kappa(U,V)d_{\alpha}(x,y,z)}}{\kappa(U,V)}\underline{m}_T(E_F^T+eU-eV)\underline{m}_S^{\alpha}(E_F^S+eU)dU\nonumber\\
\label{Eq_DT_TOPO}
D_T^{TOPO}(x,y,z,V)&=&-\varepsilon^2\frac{e^2}{h}\\
&\times&\sum_{\alpha}\int_{0}^{V}e^{-2\kappa(U,V)d_{\alpha}(x,y,z)}\frac{\partial n_T}{\partial U}(E_F^T+eU-eV)n_S^{\alpha}(E_F^S+eU)dU\nonumber\\
\label{Eq_DT_MAGN}
D_T^{MAGN}(x,y,z,V)&=&-\varepsilon^2\frac{e^2}{h}\\
&\times&\sum_{\alpha}\int_{0}^{V}e^{-2\kappa(U,V)d_{\alpha}(x,y,z)}\frac{\partial\underline{m}_T}{\partial U}(E_F^T+eU-eV)\underline{m}_S^{\alpha}(E_F^S+eU)dU.\nonumber
\end{eqnarray}
Thus, we formulated all components of the differential conductance in spin-polarized tunnel junctions within the
atom superposition framework using first principles electronic structure of the sample and the tip. Note that all $dI/dV$
expressions in Eq.(\ref{Eq_didv_all}) in appendix \ref{sec_app} can similarly be calculated within our 3D WKB approach.

$I(x,y,z,V)$ and $dI/dV(x,y,z,V)$ can be calculated at $(x,y,z)$ grid points of a three-dimensional (3D) fine grid in a
finite box above the surface. The recipe for simulating SP-STM images based on the 3D current map is given in
Ref.\ \cite{palotas11stm}. Here, we focus on the simulation of $dI/dV$ spectra.
From the 3D differential conductance map, data can be extracted that are directly comparable to experiments.
For example, a single point spectrum corresponds to a fixed $(x_0,y_0,z_0)$ tip position, and two-dimensional (2D) spectra can
also be obtained, where the image resolution is determined by the density of $(x,y)$ grid points.
There are usually two ways to define a 2D differential conductance map \cite{wiesendanger09review}.
The first method fixes the tip height at $z=Z_{stab}=const$ and scans the surface, $dI/dV(x,y,Z_{stab},V)$.
The second option measures $dI/dV$ on a constant current contour, $I_{TOTAL}=I_{stab}=const$, which is the widely
used method in experiments. Simulation of this can be done in two steps: First, we calculate the 3D current map with the given
bias voltage $V_{stab}$, and at the second step we determine the height profile of a constant current contour,
$z(x,y,V_{stab},I_{stab})$, using logarithmic interpolation \cite{palotas11stm}. $V_{stab}$ and $I_{stab}$ are the
tunneling parameters, and they stabilize the tip position at the height of $z(x,y,V_{stab},I_{TOTAL}=I_{stab})$ above the $(x,y)$
sample surface point. The 2D differential conductance map on the constant current contour is then given by
$dI/dV(x,y,z(x,y,V_{stab},I_{TOTAL}=I_{stab}),V)$, where the $V$-dependence is obtained by sweeping the bias voltage range
using a lock-in technique in experiments \cite{wiesendanger09review}.
Recently, experimental efforts have been made to extract the TOPO component of the tunneling current \cite{ding03},
and measure spectroscopic data on such constant current contours, i.e.\ at $I_{TOPO}=const$ \cite{tange10}. According to
Ref.\ \cite{wiesendanger09review}, a constant tunneling transmission enables an easier interpretation of measured 2D spectroscopic
data. We believe that a constant TOPO current contour is closer to this constant tunneling transmission criterion than a constant
TOTAL current contour due to the appearance of spin dependent effects in the latter one. On the other hand, the calculation of
{\emph{any}} current contour is simple within our 3D WKB approach \cite{palotas11stm}. Since the $I_{TOPO}=const$ experimental
method is not routinely available at the moment, we restrict ourselves to consider the $I_{TOTAL}=const$ contours when
calculating the 2D differential conductance maps, and we will show examples in the next section.

By simulating differential conductance spectra above a magnetic surface with parallel (P) and antiparallel (AP) tip magnetization
directions with respect to a pre-defined direction (usually the magnetization direction of a chosen surface atom is
taken), the so-called magnetic asymmetry can be defined \cite{zhou10}. In our case this quantity can be calculated at all
considered positions of the tip apex atom, i.e.\ at all $(x,y,z)$ grid points within our finite box above the surface:
\begin{equation}
\label{Eq_AMAGN-didv1}
A(x,y,z,V)=\frac{dI^P/dV(x,y,z,V)-dI^{AP}/dV(x,y,z,V)}{dI^P/dV(x,y,z,V)+dI^{AP}/dV(x,y,z,V)}.
\end{equation}
From this, the magnetic asymmetry can similarly be calculated on appropriate constant current contours as described
in the previous paragraph.
Using Eq.(\ref{Eq_didv_Decomp}), and the fact that the magnetic contribution for the AP tip magnetization direction
$dI_{MAGN}^{AP}/dV$ equals $-dI_{MAGN}^P/dV$, since the tip magnetization PDOS vector $\underline{m}_T(E)$ changes
sign at all energies compared to the P tip magnetization direction, the differential conductances take the following form:
\begin{eqnarray}
dI^P/dV(x,y,z,V)&=&dI_{TOPO}/dV(x,y,z,V)+dI_{MAGN}^P/dV(x,y,z,V)\nonumber\\
dI^{AP}/dV(x,y,z,V)&=&dI_{TOPO}/dV(x,y,z,V)-dI_{MAGN}^P/dV(x,y,z,V).
\end{eqnarray}
Thus, the magnetic asymmetry can be expressed as the fraction of the MAGN and TOPO differential conductances from
Eqs.\ (\ref{Eq_didvTOPO}) and (\ref{Eq_didvMAGN}) as
\begin{eqnarray}
\label{Eq_AMAGN-didv}
A^{dI/dV}(x,y,z,V)&=&\frac{dI_{MAGN}^P/dV(x,y,z,V)}{dI_{TOPO}/dV(x,y,z,V)}\\
&=&\frac{dI_{MAGN}^P/dU(x,y,z,V,V)+B_{MAGN}^P(x,y,z,V)+D_T^{MAGN,P}(x,y,z,V)}{dI_{TOPO}/dU(x,y,z,V,V)+B_{TOPO}(x,y,z,V)+D_T^{TOPO}(x,y,z,V)}.\nonumber
\end{eqnarray}
This is the correct magnetic asymmetry expression based on the physical differential conductances that can be
obtained from experiments.
However, a magnetic asymmetry can similarly be defined taking the virtual differential conductances from Eqs.\ (\ref{Eq_diduTOPO})
and (\ref{Eq_diduMAGN}):
\begin{equation}
\label{Eq_AMAGN-didu}
A^{dI/dU}(x,y,z,V)=\frac{dI_{MAGN}^P/dU(x,y,z,V,V)}{dI_{TOPO}/dU(x,y,z,V,V)}.
\end{equation}
This is an important quantity since it is related to the vacuum spin polarization of the sample in a simple way
\cite{zhou10}:
\begin{equation}
\label{Eq_AMAGN-ESP}
A^{dI/dU}(x,y,z,V)=\underline{P}_T(E_F^T)\underline{P}_S(x,y,z,E_F^S+eV)=ESP(x,y,z,V),
\end{equation}
i.e., $A^{dI/dU}(x,y,z,V)$ is the effective spin polarization (ESP): the scalar product of the tip spin polarization vector at
its Fermi level, $\underline{P}_T(E_F^T)$, and the vacuum spin polarization vector of the sample at $\underline{R}_{TIP}(x,y,z)$,
$eV$ above the sample Fermi level, $\underline{P}_S(x,y,z,E_F^S+eV)$.
Following above, it is clear that the determination of the sample spin polarization from experimentally measured spectra is not
straightforward since the experimentally accessible magnetic asymmetry according to the equivalent expressions
Eq.(\ref{Eq_AMAGN-didv1}) and Eq.(\ref{Eq_AMAGN-didv}) contains the background and tip-derivative terms as well.
On the other hand, we can easily calculate ESP$(x,y,z,V)$ within our method.
There are even more possibilities to define magnetic asymmetries, by adding the background terms in Eqs.\ (\ref{Eq_B_TOPO}) and
(\ref{Eq_B_MAGN}), or the tip-derivative terms in Eqs.\ (\ref{Eq_DT_TOPO}) and (\ref{Eq_DT_MAGN}) to the corresponding
virtual differential conductance and then performing the division:
\begin{eqnarray}
\label{Eq_AMAGN-didu_plus_B}
A^{dI/dU+B}(x,y,z,V)=\frac{dI_{MAGN}^P/dU(x,y,z,V,V)+B_{MAGN}^P(x,y,z,V)}{dI_{TOPO}/dU(x,y,z,V,V)+B_{TOPO}(x,y,z,V)},\\
A^{dI/dU+D_T}(x,y,z,V)=\frac{dI_{MAGN}^P/dU(x,y,z,V,V)+D_T^{MAGN,P}(x,y,z,V)}{dI_{TOPO}/dU(x,y,z,V,V)+D_T^{TOPO}(x,y,z,V)}.
\label{Eq_AMAGN-didu_plus_DT}
\end{eqnarray}
As $dI/dU(V,V)$ is one component of $dI/dV(V)$ according to Eq.(\ref{Eq_didv_3terms}), we will compare them and
also the magnetic asymmetry expressions in Eqs.\ (\ref{Eq_AMAGN-didv})-(\ref{Eq_AMAGN-didu_plus_DT}),
in order to estimate the error one makes when neglecting the background and tip-related components of $dI/dV(V)$
for a given combination of a complex magnetic surface and a magnetic tip. On the other hand, we will calculate
qualitatively correct bias dependent 2D effective spin polarization maps following a constant current contour.

It has to be noted that the presented method can also be applied to study nonmagnetic systems, where all magnetic
contributions equal zero and the corresponding topographic STS spectra can be simulated. Of course, in this case,
the magnetic asymmetry is zero.

\section{Results and Discussion}
\label{sec_res}

In order to demonstrate the reliability and capabilities of our model for simulating SP-STS on complex magnetic
surfaces, we consider a sample surface with noncollinear magnetic order. One ML Cr on Ag(111) is a prototype of
frustrated hexagonal antiferromagnets \cite{heinze06}. Due to the geometrical frustration of the
antiferromagnetic exchange interactions between Cr spin moments, its magnetic ground state is
a noncollinear $120^{\circ}$ N\'eel state \cite{wortmann01}. In the presence of spin-orbit coupling,
two types of N\'eel states with opposite chiralities can form, and one of them is energetically favored \cite{palotas11stm}.

We performed geometry relaxation and electronic structure calculations based on Density Functional Theory (DFT)
within the Generalized Gradient Approximation (GGA) implemented in the Vienna Ab-initio Simulation Package (VASP)
\cite{VASP2,VASP3,hafner08}. A plane wave basis set for electronic wavefunction expansion together with the
projector augmented wave (PAW) method \cite{kresse99} has been applied, while
the exchange-correlation functional is parametrized according to Perdew and Wang (PW91) \cite{pw91}.
For calculating the fully noncollinear electronic structure we used the VASP code as well \cite{hobbs00prb,hobbs00jpcm},
with spin-orbit coupling considered.

We model the Cr/Ag(111) system by a slab of a five-layer Ag substrate and one monolayer Cr film on each side,
where the surface Cr layers and the first Ag layers underneath have been fully relaxed.
A separating vacuum region of 14.6 $\AA$ width in the surface normal ($z$) direction has been set up between neighboring
supercell slabs. The average electron workfunction above the surface is $\phi_S=4.47$ eV.
We used an $11\times 11\times 1$ Monkhorst-Pack (MP) \cite{monkhorst} k-point grid for calculating the projected electron DOS
(PDOS) onto the surface Cr atoms in our ($\sqrt{3}\times\sqrt{3}$) magnetic surface unit cell \cite{palotas11stm}.

\begin{figure*}
\includegraphics[width=1.0\textwidth,angle=0]{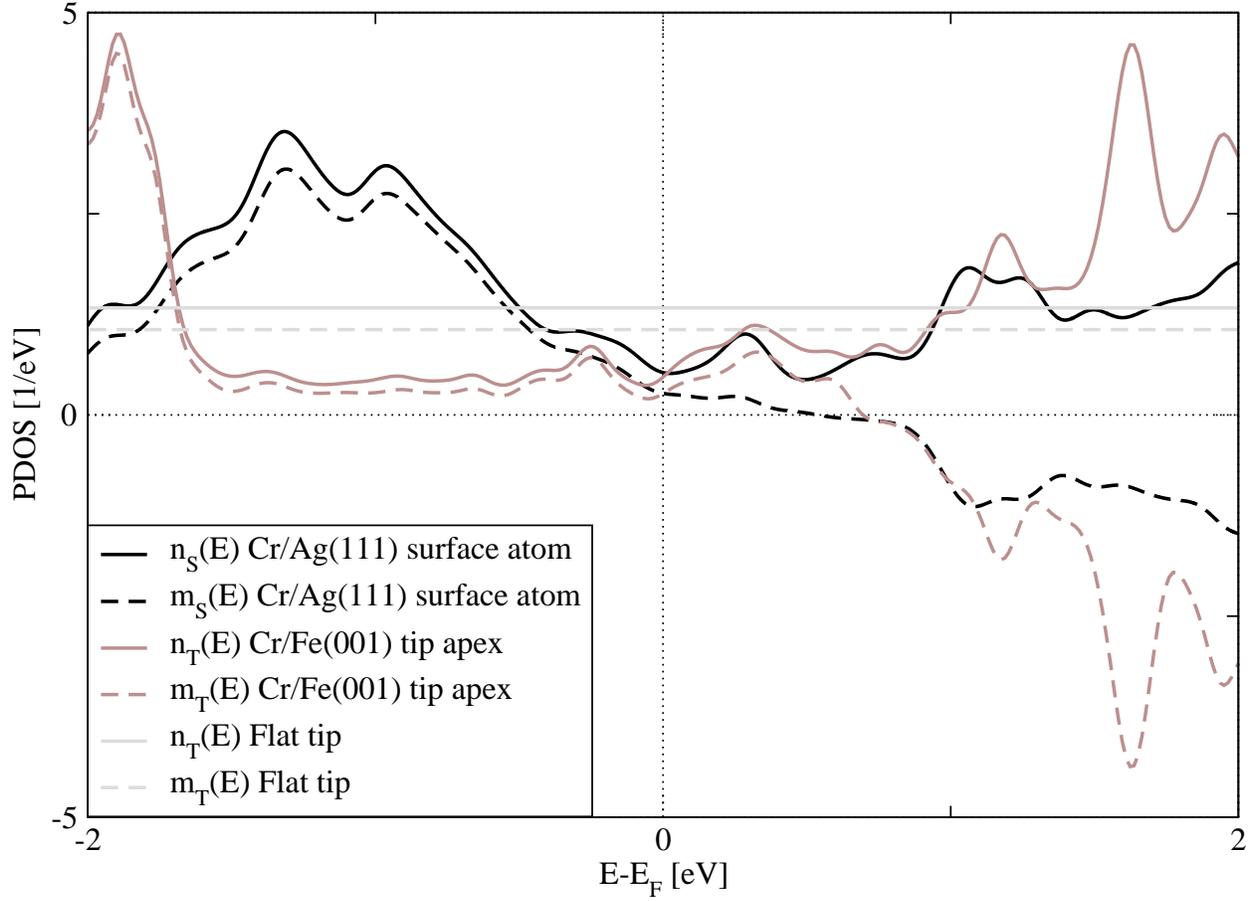}
\caption{\label{Fig1} Projected charge and magnetization DOS of the surface Cr atom of the sample Cr/Ag(111),
the tip apex Cr atom of the Cr/Fe(001) tip \cite{ferriani10tip}, and the flat tip.
}
\end{figure*}

The energy dependent charge and magnetization PDOS, $n_S(E)$ and $m_S(E)$, respectively, are shown in
Figure \ref{Fig1}. We obtained these quantities from noncollinear calculations. The spin quantization axis of each surface Cr atom
is chosen to be parallel to their magnetic moment direction, and $m_S(E)$ is the projection of the magnetization PDOS vector
$\underline{m}_S(E)$ to this direction. Except the spin quantization axes of the three different Cr atoms in the magnetic surface
unit cell, their electronic structure is the same. We can interpret the results in terms of the commonly used spin up
($\uparrow$, majority) and spin down ($\downarrow$, minority) channels with respect to the atomic spin quantization axis, where
$n(E)=n^{\uparrow}(E)+n^{\downarrow}(E)$, and $m(E)=n^{\uparrow}(E)-n^{\downarrow}(E)$.
It is seen that the majority spin PDOS dominates over the minority spin PDOS below $E_F^S+0.54$ eV, while $m_S(E)<0$ above
$E_F^S+0.54$ eV. This implies a spin polarization reversal at this particular energy \cite{palotas11stm}.

In our model, the vacuum local density of states (LDOS) is obtained by the superposition of spherically decaying
electron states according to the independent orbital approximation. Above a complex magnetic surface, the spin up and spin down
notations are meaningless since there is no global spin quantization axis. Instead, we can consider the charge and magnetization
(vector) character of the LDOS obtained from the PDOS, as defined in Eqs.\ (\ref{Eq_nLDOS}) and (\ref{Eq_mLDOS}) for $n_{LDOS}$ and
$\underline{m}_{LDOS}$, respectively. Above a surface Cr atom with lateral position $(x_0,y_0)$, both vacuum LDOS behave the same
way as the corresponding PDOS, thus the spin polarization vector in vacuum
$\underline{P}_{LDOS}(x_0,y_0,z,E)=\underline{m}_{LDOS}(x_0,y_0,z,E)/n_{LDOS}(x_0,y_0,z,E)$
equals the one obtained from the PDOS, i.e.\ $\underline{P}_S(E)=\underline{m}_S(E)/n_S(E)$. Moving out of the high symmetry
lateral position above a surface atom $(x_0,y_0)$, $\underline{m}_{LDOS}$ will vary due to the different atomic spin quantization
axes for all three Cr atoms in the magnetic surface unit cell and the considered vacuum decays. $n_{LDOS}$ will, however, remain
qualitatively unchanged. The lateral variation of $\underline{m}_{LDOS}$ will result in a position dependent
vacuum spin polarization vector of the sample surface. This quantity multiplied by the tip spin polarization vector results in the
effective spin polarization, defined in Eq.(\ref{Eq_AMAGN-ESP}), which will be simulated later.

Dependence of the tunneling spectra on the tip electronic structure can be studied by considering different tip models. In this
work we compare spectra and magnetic asymmetries measured by a magnetic CrFe tip and an electronically flat magnetic tip. The
electronic structure data of the CrFe tip apex was taken from Ref.\ \cite{ferriani10tip}, where the tip was modeled
as a single Cr apex atom on the Fe(001) surface. Ferriani et al.\ furthermore reported that an antiferromagnetic coupling of the
Cr adatom to the Fe(001) surface is energetically preferred, and the vacuum spin polarization is fairly constant at around +0.8
in the energy range $[E_F^T-1$ eV$,E_F^T+1$ eV$]$ \cite{ferriani10tip}. The local electron workfunction above the tip apex is
assumed to be $\phi_T=4.5$ eV, that has been used to obtain the energy dependent vacuum decay in Eq.(\ref{Eq_kappa_WKB}).
The charge and magnetization PDOS of the Cr apex atom, $n_T(E)$ and $m_T(E)$, respectively, are shown in
Figure \ref{Fig1}. We obtain qualitative correspondence to the PDOS of the sample Cr atom. However, due to the different
surface orientation and the different local environment of the Cr/Ag(111) surface and Cr/Fe(001) tip apex Cr atoms,
the sample and tip Cr PDOS are quantitatively different. Concerning magnetic properties, we find a spin polarization
reversal at $E_F^T+0.7$ eV. On the other hand, there is no energy dependent {\emph{vacuum}} spin polarization
reversal observed in Ref.\ \cite{ferriani10tip}. Ferriani et al.\ analyzed this effect in detail for an Fe adatom on top of
the Fe(001) surface, and they found a competition between majority $sp$ and minority $d$ states with different decays into the
vacuum. Such an orbital dependent vacuum decay is not included in our model at the moment, but work is in progress to
consider such effects.

The electronically flat magnetic tip has been modeled based on the electronic structure of the Cr apex (PDOS) of
the CrFe tip. The charge and absolute magnetization PDOS, $n_T(E)$ and $|m_T(E)|$, respectively, have been averaged
in the $[E_F^T-2$ eV$,E_F^T+2$ eV$]$ range. We obtained $n_T=1.33$/eV and $m_T=1.06$/eV, also shown in
Figure \ref{Fig1}. Thus, the spin polarization is $P_T=m_T/n_T=+0.8$. In this case, the tip-derivative term of the
differential conductance $D_T(V)$ is zero, since $\partial n_T(E)/\partial E=\partial m_T(E)/\partial E=0$.
The vacuum decay can be modeled using Eq.(\ref{Eq_kappa_WKB}), where $\kappa(U,V)$ has an explicit $V$-dependence,
and we assume that $\phi_T=\phi_S$. Alternatively, a simpler model for $\kappa(U)$ can be considered without $V$-dependence
as in Eq.(\ref{Eq_kappa_TH}).
In this case the background term of the differential conductance $B(V)$ is zero, since the tunneling transmission does not depend
on the bias voltage, and the physical differential conductance equals the virtual differential conductance,
i.e.\ $dI/dV(V)=dI/dU(V,V)$. On the other hand, by assuming a $V$-dependent vacuum decay $\kappa(U,V)$, $B(V)$ is not zero and it
contributes to the total differential conductance, i.e.\ $dI/dV(V)=dI/dU(V,V)+B(V)$.

\begin{figure*}
\includegraphics[width=1.0\textwidth,angle=0]{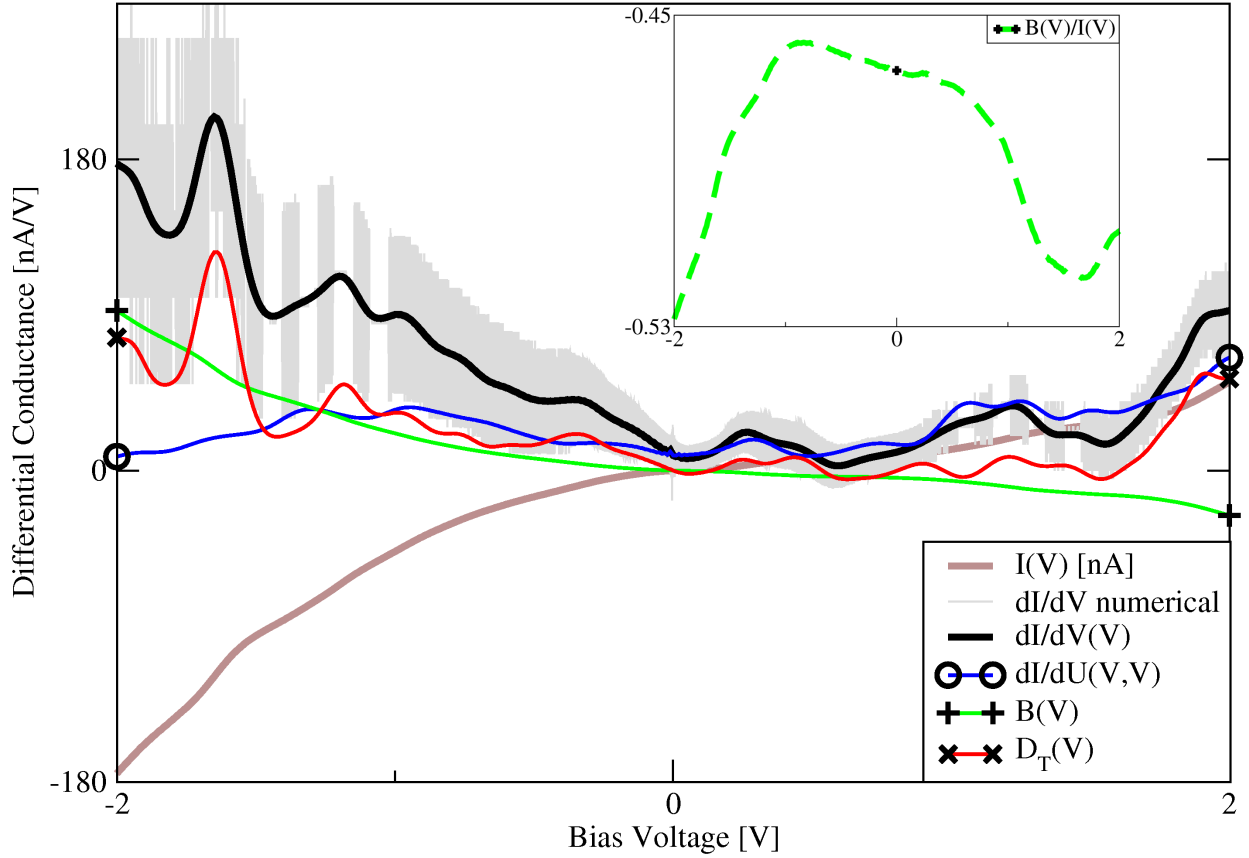}
\caption{\label{Fig2} (Color online) Comparison of single point differential conductance tunneling spectra calculated from
numerical differentiation of the tunneling current $I(V)$, and $dI/dV$ calculated according to Eq.(\ref{Eq_didv_3terms}),
and its contributing terms, the virtual differential conductance $dI/dU(V,V)$, the background term $B(V)$, and the tip-derivative
term $D_T(V)$. The model CrFe tip apex is 3.5 $\AA$ above a surface Cr atom and its magnetization direction is parallel to that of
the underlying surface Cr atom. The inset shows the ratio of $B(V)/I(V)$.
}
\end{figure*}

Figure \ref{Fig2} shows the bias dependence of the total tunneling current $I(V)$, calculated using Eq.(\ref{Eq_curr}),
at the position $z=3.5$ $\AA$ above a surface Cr atom probed with the CrFe tip having parallel (P) magnetization direction
compared to the underlying surface Cr atom.
Positive current means tunneling from the tip to the sample surface, whereas the current is negative in the opposite direction.
We find that the absolute value of the current is higher in the negative bias range compared to the positive range.
This is due to the surface and tip electronic structures. The sample occupied PDOS combined with the tip unoccupied
PDOS is greater than the sample unoccupied PDOS combined with the tip occupied PDOS, see Figure \ref{Fig1}.
Performing a numerical differentiation of $I(V)$ with respect to $V$, we obtain the differential conductance at this particular
tip position. As can be seen this is extremely noisy, and a smoothing procedure should be applied to it before further analysis.
Alternatively, the differential conductance can be calculated using Eq.(\ref{Eq_didv_3terms}), implemented within the
atom superposition approach. Figure \ref{Fig2} shows that $dI/dV$ obtained this way (black curve) is a smooth function that fits
precisely to the noisy numerical derivative of the current. There is more discussion about avoiding the
numerical differentiation of the tunneling current in determining the $dI/dV$, e.g.\ in Ref.\ \cite{hofer05sts}.

We obtain more information about the $dI/dV$ by analyzing its components, the virtual differential conductance $dI/dU(V,V)$,
the background term $B(V)$, and the tip-derivative term $D_T(V)$. We find that $dI/dU(V,V)$ differs less than
10 \% compared to $dI/dV$ in the bias range [-0.01 V, +0.01V], i.e.\ practically at the common Fermi level of tip and sample.
This means that the virtual differential conductance approximation for the $dI/dV$ (also known as the LDOS approximation) is not
sufficient except at zero bias, where they are identical, $dI/dV(0)=dI/dU(0,0)$. Moreover, one can recognize that most part of
the $dI/dV$ peak structure is already included in the $dI/dU(V,V)$ term, which is qualitatively similar to the charge
PDOS of the surface Cr atom of the sample, $n_S(E)$, see Figure \ref{Fig1}.
Apart from this, the peak structure of $D_T(V)$, calculated via Eqs.\ (\ref{Eq_DT_TOPO})-(\ref{Eq_DT_MAGN}), clearly shows up in
the $dI/dV$, particularly pronounced at high bias voltages. The reason is the rapidly changing tip electronic structure in these
energy regions, see Figure \ref{Fig1}. The features from $dI/dU(V,V)$ and $D_T(V)$
are transferred to the physical differential conductance, since $B(V)$, calculated via Eqs.\ (\ref{Eq_B_TOPO})-(\ref{Eq_B_MAGN}),
is smooth compared to the other two components in the whole bias range. Moreover, we find that $B(V)$ is a monotonous function of
the bias voltage, and it is nearly proportional to $I(V)$ as has been reported earlier for different levels of STS theories
\cite{passoni09,donati11}. The proportionality function $B(V)/I(V)$ is plotted in the inset of Figure \ref{Fig2}. It can be seen
that its sign is in agreement with Ref.\ \cite{passoni09} and it has a non-trivial bias dependence.
This is essentially due to the extra $1/\kappa(U,V)$ factor in the energy integration of the background term,
Eqs.\ (\ref{Eq_B_TOPO}) and (\ref{Eq_B_MAGN}), compared to the tunneling current expression. The $B(V)/I(V)$ function
could, in principle, be calculated at different tip-sample distances ($z$), and could be compared to analytical expressions
denoted by $f(z,V)$ reported in \cite{passoni09}. The comparison is, however, not straightforward due to two reasons. First, the
analytical expressions were reported based on the 1D WKB approximation, whereas our model is a 3D atomic superposition approach
based on WKB, which results in an effective transmission coefficient, different from the 1D WKB transmission.
Note that a 3D approach to STS with another effective transmission coefficient has recently been reported by
Donati et al.\ \cite{donati11}. Second, in Figure \ref{Fig2} we reported the sum of the TOPO and MAGN contributions,
while the related STS literature is concerned with nonmagnetic systems only, which corresponds to the analysis of the topographic
part of the spin-polarized results. Consideration of the spin-polarized tunneling complicates the analytical calculations
that are unavailable at the moment. The analysis of $B(z,V)/I(z,V)$ along the discussed lines could be a future research direction
that is beyond the scope of the present study. In the following we focus on the comparison of SP-STS spectra by reversing the
tip magnetization direction, and also using the flat magnetic tip model.

\begin{figure*}
\includegraphics[width=1.0\textwidth,angle=0]{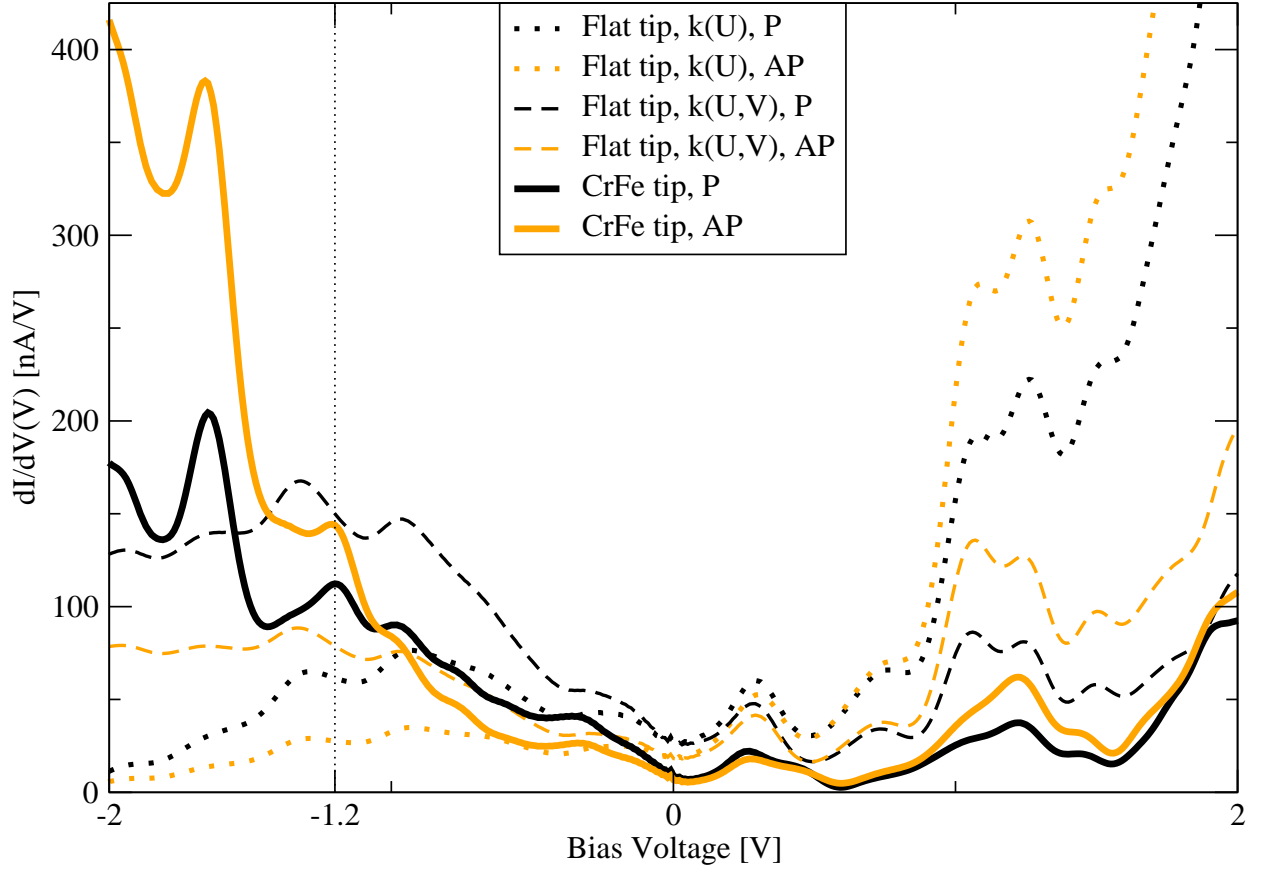}
\caption{\label{Fig3} (Color online) Comparison of simulated single point differential conductance tunneling spectra following
Eq.(\ref{Eq_didv_3terms}), probed with the flat magnetic tip and the model CrFe tip, 3.5 $\AA$ above a surface Cr atom.
Parallel (P) and antiparallel (AP) tip magnetization directions are set relative to the underneath surface Cr atom.
For the flat magnetic tip, two different vacuum decays, $\kappa(U)$ and $\kappa(U,V)$ are assumed using
Eqs.\ (\ref{Eq_kappa_TH}) and (\ref{Eq_kappa_WKB}), respectively. The vertical dotted line at -1.2 V shows the bias position of
the identified STS peak coming from the electronic structure of the CrFe tip.
}
\end{figure*}

Figure \ref{Fig3} shows simulated single point differential conductance tunneling spectra following Eq.(\ref{Eq_didv_3terms}),
probed with the flat magnetic tip and the model CrFe tip, $z=3.5$ $\AA$ above a surface Cr atom. Parallel (P) and antiparallel
(AP) tip magnetization directions are set relative to the underneath surface Cr atom. It can clearly be seen that measuring the
spectra with oppositely magnetized tips of the same type result in different differential conductance curves, in agreement with
SP-STS experiments performed on oppositely magnetized sample areas with a fixed tip magnetization direction \cite{yayon07,zhou10}.
For the flat magnetic tip, two different vacuum decays, $\kappa(U)$ and $\kappa(U,V)$ are assumed using Eqs.\ (\ref{Eq_kappa_TH})
and (\ref{Eq_kappa_WKB}), respectively. For the bias-independent vacuum decay (dotted curves) we find that $dI^P/dV>dI^{AP}/dV$
below $V=+0.54$ V, while $dI^P/dV<dI^{AP}/dV$ above $V=+0.54$ V. In our previous work \cite{palotas11sts} we identified the
effective spin polarization [$\underline{P}_T(E)\underline{P}_S(E)=\underline{m}_T(E)\underline{m}_S(E)/(n_T(E)n_S(E))$]
responsible for this effect. This is the decisive factor for
determining the sign of the magnetic contribution to $dI/dV$ at energy $E$ in the improved SP-STS model presented in section
\ref{sec_spsts} as well. The magnetic part of the physical differential conductance is given in Eq.(\ref{Eq_didvMAGN}). Since the
vacuum decay does not depend on the bias voltage $V$ for the dotted curves, and the tip is electronically flat,
$dI_{MAGN}/dV(V)=dI_{MAGN}/dU(V,V)$. Thus, the sign change of $dI_{MAGN}/dV$ occurs at the sign change of $dI_{MAGN}/dU(V,V)$,
i.e.\ at the reversal of the sample spin polarization vector at 0.54 eV above the sample Fermi level \cite{palotas11stm},
see also Figure \ref{Fig1}.
For the flat magnetic tip and the assumed bias dependent vacuum decay (dashed curves) we find that $dI^P/dV>dI^{AP}/dV$ below
$V=+0.5$ V, and $dI^P/dV<dI^{AP}/dV$ above $V=+0.5$ V, i.e.\ the sign change of the magnetic component is slightly shifted toward
zero bias. The reason is the nonzero background term $B_{MAGN}(V)$ due to $\kappa(U,V)$, and
$dI_{MAGN}/dV(V)=dI_{MAGN}/dU(V,V)+B_{MAGN}(V)$ has to be considered. Note that $D_T^{MAGN}(V)$ is still zero because of the
constant tip magnetization PDOS. Comparing the two vacuum decay models for the flat tip, it is clear that the
topographic part of the background term has another effect on the heights of the spectra, i.e.\ they are enhanced and reduced in
the negative and positive bias ranges, respectively, compared to the $\kappa(U)$ model. On the other hand, the features of the
spectra (peaks and dips) occur at the same bias positions for both vacuum decay models.

The inclusion of a realistic tip electronic structure into our model complicates the spectra even more. This is demonstrated in
Figure \ref{Fig3} for the CrFe tip model (solid lines). In this case all three terms contribute to the differential conductance,
and $dI/dV(V)=dI/dU(V,V)+B(V)+D_T(V)$. Thus, the relative heights of the differential conductance tunneling spectra $dI^P/dV$ and
$dI^{AP}/dV$ are determined by the superposition of the magnetic $dI_{MAGN}/dU(V,V)$, $B_{MAGN}(V)$, and $D_T^{MAGN}(V)$ terms.
The role of the effective spin polarization is more complicated, since, apart from the $dI_{MAGN}/dU(V,V)$ term, it appears in the
$dI/dV$ expression through the bias-integrated quantities $B_{MAGN}(V)$ and $D_T^{MAGN}(V)$.
For the P tip magnetization, $dI^P/dV$ is the same as the black solid curve in Figure \ref{Fig2}, and its contributions are also
shown there. In Figure \ref{Fig3}, we observe more changes of the relative height of the $dI^P/dV$ and $dI^{AP}/dV$ spectra
measured with the CrFe tip than with the flat tip. These include the sign changes of the magnetic part of the spectra, similarly
as before. We find that $dI^P/dV>dI^{AP}/dV$ in the bias interval [-1.04 V, +0.49 V], and a reversed relation is obtained in the
complementary bias regime. Comparing the spectra to the ones measured with the flat magnetic tip, we see that they are
qualitatively closer to the $\kappa(U,V)$ model used for the flat tip due to the presence of the background terms.
Moreover, the individual features coming from the sample and the tip electronic structures can be assigned.
In our case we identify the peak at -1.2 V, indicated by a vertical dotted line in Figure \ref{Fig3}, coming from the CrFe tip
electronic structure since it is missing from the spectra calculated with the flat tip. All other features are
related to the sample electronic structure as they appear in the spectra measured with the flat tip.

\begin{figure*}
\includegraphics[width=1.0\textwidth,angle=0]{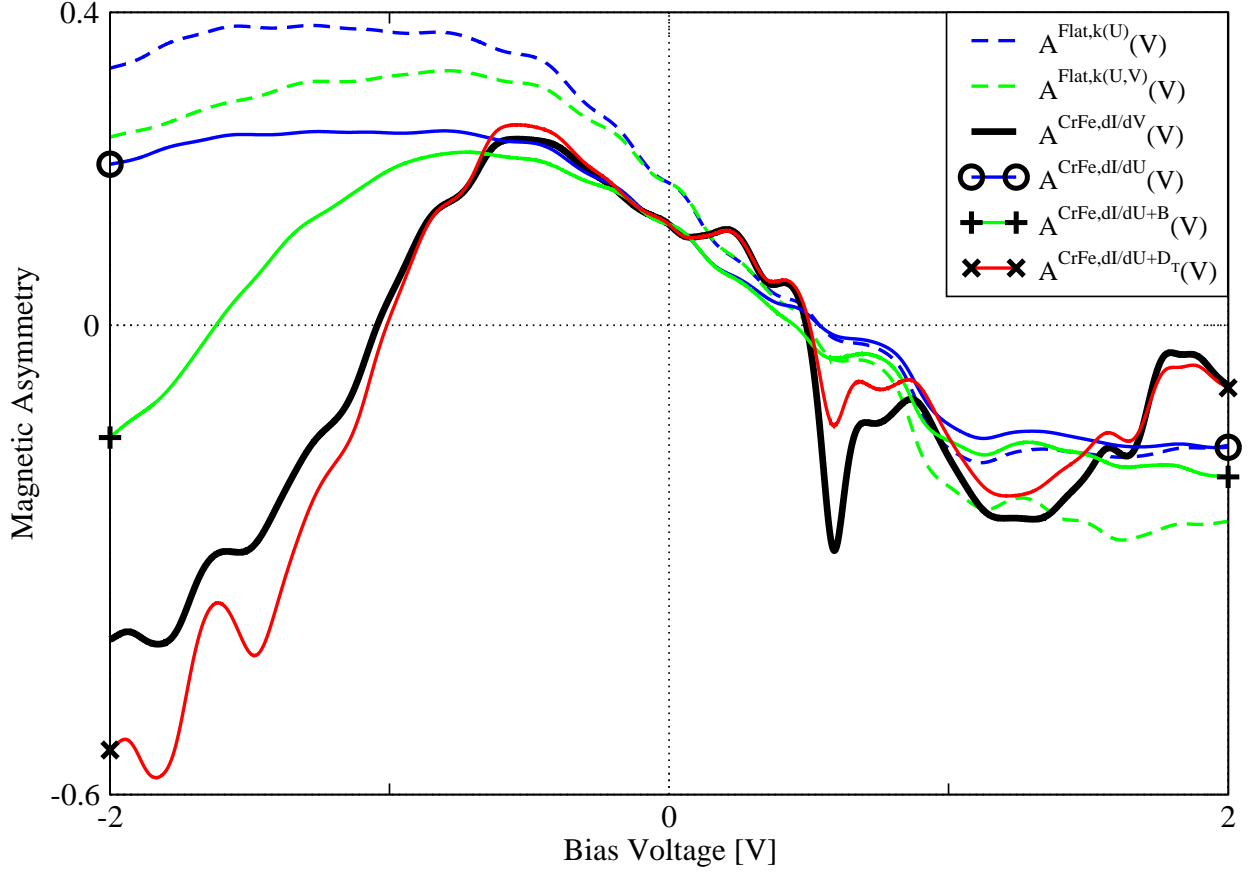}
\caption{\label{Fig4} (Color online) Comparison of magnetic asymmetries 3.5 $\AA$ above a surface Cr atom probed with the
flat magnetic tip and the model CrFe tip. $A^{Flat,\kappa(U)}$, $A^{Flat,\kappa(U,V)}$, and $A^{CrFe,dI/dV}$ are calculated from
the corresponding P and AP spectra shown in Figure \ref{Fig3}. For the CrFe tip we compare the magnetic asymmetry expressions
defined in Eqs.\ (\ref{Eq_AMAGN-didv})-(\ref{Eq_AMAGN-didu_plus_DT}).
}
\end{figure*}

The relative heights of the differential conductance tunneling spectra $dI^P/dV$ and $dI^{AP}/dV$ can also be determined from
the magnetic asymmetry, Eq.(\ref{Eq_AMAGN-didv1}). Let us compare the magnetic asymmetries calculated from the spectra in Figure
\ref{Fig3} using the two magnetic tips. Moreover, for the CrFe tip we compare the asymmetry expressions defined in
Eqs.\ (\ref{Eq_AMAGN-didv})-(\ref{Eq_AMAGN-didu_plus_DT}), in order to estimate the error one makes when neglecting the
background and tip-related components of $dI/dV(V)$. Figure \ref{Fig4} shows the calculated asymmetry functions at $z=3.5\AA$
above a surface Cr atom. It can be seen that $A^{Flat,\kappa(U)}(V)$ and $A^{Flat,\kappa(U,V)}(V)$ (dashed curves)
behave qualitatively similarly. In addition, $A^{Flat,\kappa(U)}(V)$ is greater than
$A^{Flat,\kappa(U,V)}(V)$ in almost the full studied bias range. The opposite relation holds between 0 V and +0.3 V only,
however, the relative difference between the two quantities is less than 1.4 \% in this regime. Moreover, these two magnetic
asymmetries are within 5\% relative difference in the bias range [-0.23 V, +0.31 V].

Considering the CrFe tip, the experimentally measurable magnetic asymmetry $A^{CrFe,dI/dV}(V)$ (black solid curve)
is qualitatively different from the two asymmetry functions calculated with the flat tip, e.g.\ it has a richer structure at
positive bias voltages. More importantly, it has an extra sign change occurring at -1.04 V apart from +0.49 V.
These correspond to the height changes of $dI^P/dV$ and $dI^{AP}/dV$ relative to each other in Figure \ref{Fig3}.
Let us estimate the error of the magnetic asymmetry when neglecting the background and the tip-derivative terms.
According to Eq.(\ref{Eq_AMAGN-didu}), $A^{CrFe,dI/dU}(V)$ (curve with symbol 'o') considers the virtual differential
conductances only. It is within 10\% relative error compared to $A^{CrFe,dI/dV}(V)$ in the bias range [-0.65 V, +0.1 V].
However, its sign does not correspond to $A^{CrFe,dI/dV}(V)$ in the bias intervals [-2 V, -1.04 V] and [+0.49 V, +0.54 V].
Adding the background term $B(V)$ to $dI/dU(V,V)$ results in an improved differential conductance expression, and
$A^{CrFe,dI/dU+B}(V)$ (curve with symbol '+'), defined in Eq.(\ref{Eq_AMAGN-didu_plus_B}), behaves qualitatively similarly
to $A^{CrFe,dI/dU}(V)$ above -0.65 V. However, its sign change is shifted to +0.45 V from +0.54 V.
Additionally, a sign change in the negative bias range occurs at -1.62 V.
Close to the sample Fermi level, $A^{CrFe,dI/dU+B}(V)$ is within 10\% relative error compared to
$A^{CrFe,dI/dV}(V)$ in a decreased bias range of [-0.34 V, +0.1 V].
Finally, by adding the tip-derivative term $D_T(V)$ to $dI/dU(V,V)$, $A^{CrFe,dI/dU+D_T}(V)$ (curve with symbol 'x'),
defined in Eq.(\ref{Eq_AMAGN-didu_plus_DT}), shows the most closely related shape to $A^{CrFe,dI/dV}(V)$. Furthermore,
it is also quantitatively close to the physical magnetic asymmetry as its sign changes occur at -1.01 V and +0.5 V, and
it is within 10\% relative error compared to $A^{CrFe,dI/dV}(V)$ in an increased bias interval [-0.90 V, +0.45 V].
Summarizing this paragraph, the contribution of all three terms to the $dI/dV(V)$ according to Eq.(\ref{Eq_didv_3terms}) is
needed to define the physical magnetic asymmetry, that should be comparable to experiments.

\begin{figure*}
\includegraphics[width=0.8\textwidth,angle=0]{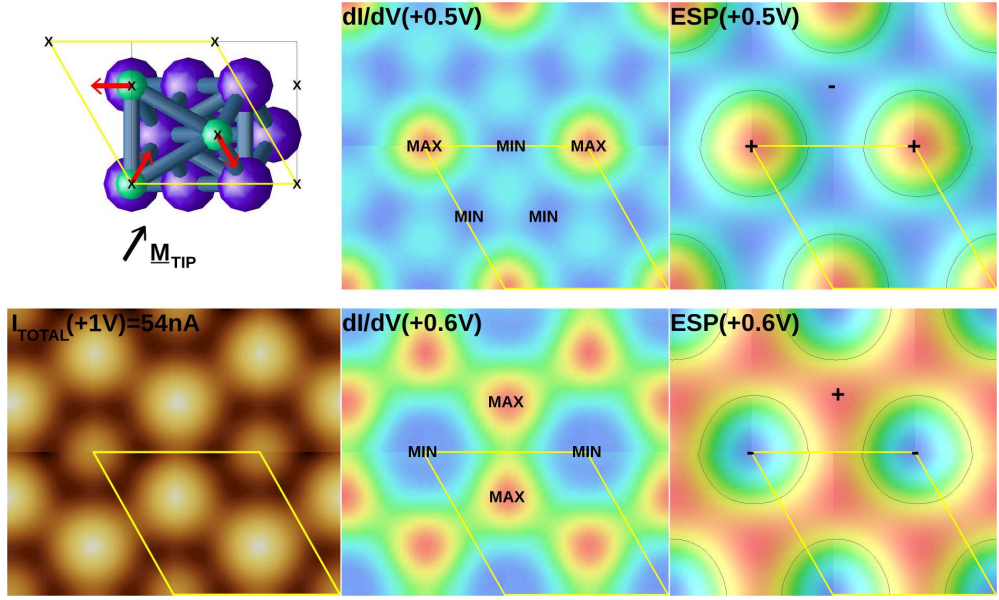}
\caption{\label{Fig5} (Color online) Top left: Surface geometry of 1 ML Cr on Ag(111). The Cr and Ag atoms are
denoted by spheres colored by green (medium gray) and purple (dark gray), respectively, while the magnetic moments of individual
Cr atoms are indicated by (red) arrows. The ($\sqrt{3}\times\sqrt{3}$) magnetic unit cell is drawn by yellow (light gray) color.
The surface Cr positions are denoted by 'x'. Bottom left: Constant current contour about 3.5 $\AA$ above the surface with
$I_{TOTAL}(V_{stab}=+1$V$)$=54 nA calculated with the flat magnetic tip using $\kappa(U,V)$, Eq.(\ref{Eq_kappa_WKB}).
The tip magnetization direction ($\underline{M}_{TIP}$) is indicated by an arrow. Middle column: Simulated 2D differential
conductance maps $dI/dV(x,y,V=+0.5$V$)$ (top middle; min. 15.8, max. 16.5 nA/V), and $dI/dV(x,y,V=+0.6$V$)$ (bottom middle;
min. 22.9, max. 23.6 nA/V), while the tip is following the constant current contour at the bottom left of the figure.
Minimum (MIN) and maximum (MAX) values are indicated. Right column: Simulated effective spin polarization (ESP) maps on the
same current contour following Eq.(\ref{Eq_AMAGN-ESP}), ESP$(x,y,V=+0.5$V$)$ (top right), and ESP$(x,y,V=+0.6$V$)$ (bottom right).
Black contours correspond to zero ESP, and the regions with positive (+) and negative (-) ESP are indicated.
The surface magnetic unit cell is drawn by a yellow (light gray) rhombus on each 2D map.
}
\end{figure*}

Our method presented in section \ref{sec_spsts} also enables one to simulate two-dimensional (2D) $dI/dV$ and magnetic asymmetry
maps in high spatial resolution above the surface, that can be compared to results of SP-STS experiments.
Such experiments are routinely performed while the tip follows a constant TOTAL current contour,
see e.g.\ Ref.\ \cite{kubetzka05}.
Figure \ref{Fig5} illustrates this capability of our method, where we used the flat tip model with tip magnetization
direction $\underline{M}_{TIP}$ parallel to the $(1/2,\sqrt{3}/2)$ direction (i.e.\ the magnetization direction of the surface Cr
atom at the bottom left corner of the scan area, see top left part of Figure \ref{Fig5}). Moreover, $\kappa(U,V)$,
Eq.(\ref{Eq_kappa_WKB}) has been used for the vacuum decay. By choosing $V_{stab}=+1$ V,
we calculate the 3D TOTAL current map in a box above the surface. From this 3D data we extract the
current contour of $I_{TOTAL}=54$ nA, which is around 3.5 $\AA$ above the sample surface and has a corrugation of 4.2 pm.
This contour, $z(x,y,V_{stab}=+1$V$,I_{TOTAL}=54$nA$)$ is plotted in the bottom left part of
Figure \ref{Fig5}. The apparent height of the Cr atom with parallel magnetic moment to the tip is lower than those of the other
two Cr atoms in the magnetic surface unit cell. This has been explained in a previous work \cite{palotas11stm}.
The surface scan area and the magnetic unit cell are shown in the top left part of Figure \ref{Fig5},
indicated by a black-bordered rectangle, and a yellow (light gray) rhombus, respectively.
For calculating the differential conductance-related 2D maps, we vary the vertical position
$z$ of the tip apex atom following the constant current contour shown in the bottom left part of Figure \ref{Fig5}.
Thus, spin-resolved $dI/dV$ and magnetic asymmetry maps can be simulated at different $V$ bias voltages corresponding to
experiments. As an example, $dI/dV(x,y)$ and the effective spin polarization ESP$(x,y)$, see Eq.(\ref{Eq_AMAGN-ESP}), are shown
in the middle and right columns of Figure \ref{Fig5}, respectively, calculated at bias voltages
$V=+0.5$ V (top) and $V=+0.6$ V (bottom). We chose these voltages close to the spin polarization reversal of the sample surface at
0.54 eV above its Fermi level, see Figure \ref{Fig1}, and Ref.\ \cite{palotas11stm}. Indeed, the reversal of the 2D $dI/dV$ map
at $V=+0.6$ V compared to $V=+0.5$ V can clearly be seen. While the SP-STM image at +1 V and the $dI/dV$ map at +0.6 V show the
same type of contrast, the $dI/dV$ signal is inverted for +0.5 V. Since $P_T=+0.8$ is constant in the full energy range, this
effect is due to the surface electronic structure. At +0.6 V bias, all surface Cr spin polarization vectors point
{\emph{opposite}} to their local magnetic moment directions \cite{palotas11stm}, and since $P_T=+0.8$ is set with
respect to the $(1/2,\sqrt{3}/2)$ direction ($\underline{M}_{TIP}$), the leading term of
the magnetic differential conductance, $dI_{MAGN}/dU(V,V)$ is negative above the surface Cr atom with parallel
magnetic moment to the tip. Moreover, the sign of $dI_{MAGN}/dU(V,V)$ changes to positive above the other two Cr atoms in the
magnetic unit cell. This results in the minimal total $dI/dV(x,y)$ above the Cr atom at the bottom left corner of
the scan area (22.9 nA/V, magn.\ moment parallel to tip), whereas above the other two Cr atoms $dI/dV$ is maximal
(23.6 nA/V, magn.\ moment not in line with tip). This happens even though the topographic differential
conductance is higher above the Cr atom which is lower-lying on the constant current contour. Similarly, the case of +0.5 V is
reversed, since all surface Cr spin polarization vectors point {\emph{along}} their local magnetic moment directions
\cite{palotas11stm} and the maximal total $dI/dV(x,y)$ is achieved above the Cr atom at the bottom left corner of
the scan area (16.5 nA/V, magn.\ moment parallel to tip), whereas above the other two Cr atoms $dI/dV$ is lower
(16.0 nA/V, magn.\ moment not in line with tip). The minimal $dI/dV$=15.8 nA/V is obtained above the midpoint of
the lines connecting two $dI/dV$ maxima. If we introduce the notation of $dI^P/dV(x,y)$ for the above
calculated differential conductances with P parallel to the indicated $\underline{M}_{TIP}$ in Figure \ref{Fig5}, then the
antiparallel tip orientation is denoted by AP, and $dI^{AP}/dV(x,y)$ can similarly be calculated.
For the very same reason as discussed, a reversed tip magnetization direction would result in a reversed
$dI^{AP}/dV$ map concerning the heights above the non-equivalent magnetic Cr atoms.
Thus, at +0.6 V the difference between $dI^P/dV(x,y)$ and $dI^{AP}/dV(x,y)$
is minimal and negative above the bottom left Cr atom in the scan area, and maximal and positive above the other two Cr atoms,
while the opposite is true at +0.5 V.
These explain qualitatively well the simulated ESP$(x,y)$ maps, see the right column of Figure \ref{Fig5}.
The ESP$(x,y)=0$ contour acts as a border between surface regions with positive and negative ESP at the given bias. Note that the
sign of the tip spin polarization has a crucial effect on the ESP$(x,y)$ map. Reversing the sign of $P_T$ compared
to the $\underline{M}_{TIP}$ direction would result in a reversed ESP$(x,y)$ map.

We suggest that by applying our method to magnetic surfaces, two-dimensional $dI^P/dV(x,y)$, $dI^{AP}/dV(x,y)$, and
magnetic asymmetry $A(x,y)$ maps can be constructed on appropriate current contours at arbitrary $V$ bias,
corresponding to SP-STS experiments. Similarly, an ESP$(x,y)$ map can be simulated. We stress again that the ESP can not simply be
obtained from experimental magnetic asymmetry due to the presence of the background and tip-derivative terms.
By explicitly considering the tip electronic structure in our SP-STS model based on experimental information,
it would help in a more reasonable interpretation of experimentally measured tunneling spectra, magnetic asymmetries,
and effective spin polarization.

\section{Conclusions}
\label{sec_conc}

We presented an efficient simulation method for spin-polarized scanning tunneling spectroscopy based on
first principles electronic structure data within our atom superposition framework \cite{palotas11stm} by including
the bias dependent background and tip-derivative terms into the differential conductance formula
following Passoni et al.\ \cite{passoni09}. We showed that our simulated data can be related to standard experimental setups.
Taking the tip electronic structure into account, the effect of a richer variety of electronic structure properties can be
investigated on the tunneling transport within the indicated approximations
(atom superposition, orbital-independent spherical vacuum decay).
The method is computationally cheap and it can be applied based on the results of any ab initio electronic structure code.
Taking a prototype frustrated hexagonal antiferromagnetic system, a Cr monolayer on Ag(111) in a noncollinear magnetic
$120^{\circ}$ N\'eel state, we simulated differential conductance tunneling spectra and magnetic asymmetries to illustrate the
applicability of our method, and we analyzed the contributing terms. We found that the features of the tunneling spectra are
coming from the virtual differential conductance and tip-derivative terms, and the background term is proportional to the
tunneling current. We showed evidence that the tunneling spectra and the related magnetic asymmetries are sensitive to the
tip electronic structure and to the vacuum decay. We also demonstrated a simulation method for 2D $dI/dV$,
magnetic asymmetry, and qualitatively correct effective spin polarization maps above a complex magnetic surface
following a constant current contour. Finally, we pointed out that the magnetic asymmetry obtained from experiments
can not simply be related to the sample spin polarization due to the presence of the background and tip-derivative terms.

\section*{Acknowledgments}

The authors thank Paolo Ferriani and Stefan Heinze for providing the electronic structure data of the CrFe tip.
Financial support of the Magyary Foundation, EEA and Norway Grants, the Hungarian Scientific Research Fund (OTKA PD83353, K77771),
the Bolyai Research Grant of the Hungarian Academy of Sciences,
and the New Sz\'echenyi Plan of Hungary (Project ID: T\'AMOP-4.2.2.B-10/1--2010-0009) is gratefully acknowledged.

\appendix
\section{Theory of STS within 1D WKB}
\label{sec_app}

We report the formulation of the tunneling current and the differential conductance in the framework of the
one-dimensional (1D) WKB approximation, which has been used in our atom superposition approach in section \ref{sec_spsts}.
Assuming elastic tunneling, the non-spinpolarized part of the tunneling current at zero temperature is given by
\cite{ukraintsev96,passoni07}
\begin{equation}
I(V,d)=C\int_{E_F^S}^{E_F^S+eV}T(E,V,d)n_T(E)n_S(E)dE,
\end{equation}
where $V$ is the bias voltage, $d$ the tip-sample distance, $C$ an appropriate constant, $E_F^S$ the Fermi energy of the sample
surface, $e$ the elementary charge, $T$ the tunneling transmission coefficient, while $n_T(E)$ and $n_S(E)$ are the tip and sample
densities of states, respectively. Performing a change of variable from $E$ to $U$ using $E=E_F^S+eU$, the tunneling current reads
\begin{equation}
I(V,d)=Ce\int_{0}^{V}T(E_F^S+eU,V,d)n_T(E_F^S+eU)n_S(E_F^S+eU)dU.
\end{equation}
The applied bias voltage $V$ in the tunnel junction defines the difference between tip and sample Fermi levels,
$E_F^T=E_F^S+eV$. Using this, the energy dependence of $n_T(E)$ can be rewritten related to the tip Fermi level $E_F^T$, and
the tunneling current can be reformulated as
\begin{equation}
I(V,d)=Ce\int_{0}^{V}T(E_F^S+eU,V,d)n_T(E_F^T+eU-eV)n_S(E_F^S+eU)dU.
\end{equation}
We denote the integrand by the formal quantity,
\begin{equation}
\label{Eq_Integrand}
\frac{dI}{dU}(U,V,d)=CeT(E_F^S+eU,V,d)n_T(E_F^T+eU-eV)n_S(E_F^S+eU),
\end{equation}
called virtual differential conductance. The tunneling current can then be expressed as
\begin{equation}
\label{Eq_I_formal}
I(V,d)=\int_{0}^{V}\frac{dI}{dU}(U,V,d)dU.
\end{equation}
The physical differential conductance can be obtained as the derivative of the tunneling current with respect to the
bias voltage. This can formally be written as
\begin{equation}
\label{Eq_dIdV-formal}
\frac{dI}{dV}(V,d)=\frac{dI}{dU}(V,V,d)+\int_{0}^{V}\left.\frac{\partial}{\partial V'}\frac{dI}{dU}(U,V',d)\right|_{V'=V}dU,
\end{equation}
or using Eq.(\ref{Eq_Integrand}) as
\begin{eqnarray}
\label{Eq_dIdV-proper}
\frac{dI}{dV}(V,d)&=&CeT(E_F^S+eV,V,d)n_T(E_F^T)n_S(E_F^S+eV)\\
&+&Ce\int_{0}^{V}\left.\frac{\partial}{\partial V'}\left[T(E_F^S+eU,V',d)n_T(E_F^T+eU-eV')\right]\right|_{V'=V}n_S(E_F^S+eU)dU.\nonumber
\end{eqnarray}
This is a known formula in the literature \cite{ukraintsev96,passoni07}.
If the tip electronic structure is assumed to be energetically flat, i.e.\ $n_T(E)=n_T$, which is still a widely used
approximation in the recent literature, then the $V$-dependence of $n_T(E_F^T+eU-eV)$ disappears, i.e.\ $n_T(E_F^T+eU-eV)=n_T$,
and the differential conductance becomes
\begin{eqnarray}
\frac{dI}{dV}(V,d)&=&CeT(E_F^S+eV,V,d)n_T(E_F^T)n_S(E_F^S+eV)\\
&+&Cen_T\int_{0}^{V}\frac{\partial T}{\partial V}(E_F^S+eU,V,d)n_S(E_F^S+eU)dU\nonumber
\end{eqnarray}
Here, the second term is the so-called background term, which is a monotonous function of the bias voltage \cite{passoni07}.
Going beyond the assumption of the electronically flat tip by incorporating the tip electronic structure in the
differential conductance expression, the effect of the tip can be studied on the tunneling spectra.
The explicit energy dependence of $n_T(E)$ can be calculated from first principles \cite{ferriani10tip,palotas11sts}, or can be
included in a model way \cite{passoni09}.
Following Eq.(\ref{Eq_dIdV-proper}), the differential conductance can be reformulated as
\begin{eqnarray}
\frac{dI}{dV}(V,d)&=&CeT(E_F^S+eV,V,d)n_T(E_F^T)n_S(E_F^S+eV)\\
&+&Ce\int_{0}^{V}\frac{\partial T}{\partial V}(E_F^S+eU,V,d)n_T(E_F^T+eU-eV)n_S(E_F^S+eU)dU\nonumber\\
&+&Ce\int_{0}^{V}T(E_F^S+eU,V,d)\frac{\partial n_T}{\partial V}(E_F^T+eU-eV)n_S(E_F^S+eU)dU.\nonumber
\end{eqnarray}
Using that $\partial n_T(E_F^T+eU-eV)/\partial V=-\partial n_T(E_F^T+eU-eV)/\partial U$,
the differential conductance at bias voltage $V$ can be written as a sum of three terms,
\begin{equation}
\label{Eq_didv2}
\frac{dI}{dV}(V,d)=\frac{dI}{dU}(V,V,d)+B(V,d)+D_T(V,d)
\end{equation}
with
\begin{eqnarray}
\frac{dI}{dU}(V,V,d)&=&CeT(E_F^S+eV,V,d)n_T(E_F^T)n_S(E_F^S+eV)\\
B(V,d)&=&Ce\int_{0}^{V}\frac{\partial T}{\partial V}(E_F^S+eU,V,d)n_T(E_F^T+eU-eV)n_S(E_F^S+eU)dU\label{Eq_B}\\
D_T(V,d)&=&-Ce\int_{0}^{V}T(E_F^S+eU,V,d)\frac{\partial n_T}{\partial U}(E_F^T+eU-eV)n_S(E_F^S+eU)dU.\label{Eq_DT}
\end{eqnarray}
Here, $B(V,d)$ is the background term usually considered in recent STS theories \cite{passoni07,passoni09,donati11},
and $D_T(V,d)$ is a term containing the energy derivative of the tip density of states (DOS), which is rarely
taken into account for practical STS calculations and analyses of experimental STS data.

It can be shown that an alternative expression for the differential conductance can be derived using integration by parts,
\begin{equation}
\label{Eq_didv1}
\frac{dI}{dV}(V,d)=\frac{dI}{dU}(0,V,d)+B(V,d)+B_2(V,d)-D_S(V,d)
\end{equation}
with
\begin{eqnarray}
\frac{dI}{dU}(0,V,d)&=&CeT(E_F^S,V,d)n_T(E_F^T-eV)n_S(E_F^S)\\
B_2(V,d)&=&Ce\int_{0}^{V}\frac{\partial T}{\partial U}(E_F^S+eU,V,d)n_T(E_F^T+eU-eV)n_S(E_F^S+eU)dU\label{Eq_B2}\\
D_S(V,d)&=&-Ce\int_{0}^{V}T(E_F^S+eU,V,d)n_T(E_F^T+eU-eV)\frac{\partial n_S}{\partial U}(E_F^S+eU)dU.\label{Eq_DS}
\end{eqnarray}
This way another background term, $B_2(V,d)$ enters the differential conductance formula, and the energy derivative of the
sample DOS appears in the term $D_S(V,d)$. The average of the two $dI/dV$ expressions can also be formed as
\begin{equation}
\label{Eq_didv3}
\frac{dI}{dV}(V,d)=\frac{1}{2}\left[\frac{dI}{dU}(0,V,d)+\frac{dI}{dU}(V,V,d)\right]+B(V,d)+\frac{1}{2}B_2(V,d)+\frac{1}{2}\left[D_T(V,d)-D_S(V,d)\right],
\end{equation}
which gives a third alternative form for the differential conductance within the 1D WKB approximation.
On the other hand, by subtracting Eq.(\ref{Eq_didv1}) from Eq.(\ref{Eq_didv2}), one gets
\begin{equation}
0=\left[\frac{dI}{dU}(V,V,d)-\frac{dI}{dU}(0,V,d)\right]-\left[B_2(V,d)-D_T(V,d)-D_S(V,d)\right].
\end{equation}
This is trivial since $B_2(V,d)-D_T(V,d)-D_S(V,d)$ is related to the partial derivative of $dI/dU(U,V,d)$ with respect to $U$:
\begin{equation}
B_2(V,d)-D_T(V,d)-D_S(V,d)=Ce\int_{0}^{V}\frac{\partial}{\partial U}\frac{dI}{dU}(U,V,d)dU=\frac{dI}{dU}(V,V,d)-\frac{dI}{dU}(0,V,d).
\end{equation}
From the three equivalent $dI/dV$ formulas in Eqs.\ (\ref{Eq_didv2}), (\ref{Eq_didv1}) and (\ref{Eq_didv3}), the calculation of
Eq.(\ref{Eq_didv2}) needs the least mathematical operations, thus, we adopted this formula to our atom superposition approach
in section \ref{sec_spsts} in order to simulate STS spectra based on electronic structure data calculated from first
principles.

Finally, note that using the transmission function in Eq.(\ref{Eq_Transmission}), and the given form of the
vacuum decay in Eq.(\ref{Eq_kappa_WKB}), the derivative of the transmission probability with respect to $U$ is obtained as
\begin{equation}
\frac{\partial T}{\partial U}(E_F^S+eU,V,d)=2med\frac{T(E_F^S+eU,V,d)}{\hbar^2\kappa(U,V)}=-2\frac{\partial T}{\partial V}(E_F^S+eU,V,d).
\end{equation}
Here, we also considered the bias-derivative of the transmission, Eq.(\ref{Eq_TransmissionDer}).
Therefore, for this particular transmission function, $B_2(V,d)=-2B(V,d)$, and the $dI/dV$ can be expressed as
\begin{eqnarray}
\label{Eq_didv_all}
\frac{dI}{dV}(V,d)&=&\frac{dI}{dU}(V,V,d)+B(V,d)+D_T(V,d)\\
&=&\frac{dI}{dU}(0,V,d)-B(V,d)-D_S(V,d)\nonumber\\
&=&\frac{1}{2}\left[\frac{dI}{dU}(0,V,d)+\frac{dI}{dU}(V,V,d)\right]+\frac{1}{2}[D_T(V,d)-D_S(V,d)].\nonumber
\end{eqnarray}
This formulation helps the better understanding of the structure of the differential conductance, and its contributing terms, and
could prove to be useful for extracting information about the tip and sample electronic structures from experimentally measured
spectra in the future.

\newpage

\begin{thebibliography}{99}

\expandafter\ifx\csname natexlab\endcsname\relax\def\natexlab#1{#1}\fi
\expandafter\ifx\csname bibnamefont\endcsname\relax
  \def\bibnamefont#1{#1}\fi
\expandafter\ifx\csname bibfnamefont\endcsname\relax
  \def\bibfnamefont#1{#1}\fi
\expandafter\ifx\csname citenamefont\endcsname\relax
  \def\citenamefont#1{#1}\fi
\expandafter\ifx\csname url\endcsname\relax
  \def\url#1{\texttt{#1}}\fi
\expandafter\ifx\csname urlprefix\endcsname\relax\def\urlprefix{URL }\fi
\providecommand{\bibinfo}[2]{#2}
\providecommand{\eprint}[2][]{\url{#2}}

\bibitem[{\citenamefont{Binnig et~al.}(1982)\citenamefont{Binnig, Rohrer,
  Gerber, and Weibel}}]{binnig1}
\bibinfo{author}{\bibfnamefont{G.}~\bibnamefont{Binnig}},
  \bibinfo{author}{\bibfnamefont{H.}~\bibnamefont{Rohrer}},
  \bibinfo{author}{\bibfnamefont{C.}~\bibnamefont{Gerber}}, \bibnamefont{and}
  \bibinfo{author}{\bibfnamefont{E.}~\bibnamefont{Weibel}},
  \bibinfo{journal}{Appl. Phys. Lett.} \textbf{\bibinfo{volume}{40}},
  \bibinfo{pages}{178} (\bibinfo{year}{1982}).

\bibitem[{\citenamefont{Binnig et~al.}(1982)\citenamefont{Binnig, Rohrer,
  Gerber, and Weibel}}]{binnig2}
\bibinfo{author}{\bibfnamefont{G.}~\bibnamefont{Binnig}},
  \bibinfo{author}{\bibfnamefont{H.}~\bibnamefont{Rohrer}},
  \bibinfo{author}{\bibfnamefont{C.}~\bibnamefont{Gerber}}, \bibnamefont{and}
  \bibinfo{author}{\bibfnamefont{E.}~\bibnamefont{Weibel}},
  \bibinfo{journal}{Phys. Rev. Lett.} \textbf{\bibinfo{volume}{49}},
  \bibinfo{pages}{57} (\bibinfo{year}{1982}).

\bibitem[{\citenamefont{Hofer et~al.}(2003)\citenamefont{Hofer, Foster,
  and Shluger}}]{hofer03rmp}
\bibinfo{author}{\bibfnamefont{W.~A.}~\bibnamefont{Hofer}},
  \bibinfo{author}{\bibfnamefont{A.~S.}~\bibnamefont{Foster}}, \bibnamefont{and}
  \bibinfo{author}{\bibfnamefont{A.~L.}~\bibnamefont{Shluger}},
  \bibinfo{journal}{Rev. Mod. Phys.} \textbf{\bibinfo{volume}{75}},
  \bibinfo{pages}{1287} (\bibinfo{year}{2003}).

\bibitem[{\citenamefont{Hofer}(2003)}]{hofer03pssci}
\bibinfo{author}{\bibfnamefont{W.~A.}~\bibnamefont{Hofer}},
  \bibinfo{journal}{Prog. Surf. Sci.} \textbf{\bibinfo{volume}{71}},
  \bibinfo{pages}{147} (\bibinfo{year}{2003}).

\bibitem[{\citenamefont{Ukraintsev}(1996)}]{ukraintsev96}
\bibinfo{author}{\bibfnamefont{V.~A.}~\bibnamefont{Ukraintsev}},
  \bibinfo{journal}{Phys. Rev. B} \textbf{\bibinfo{volume}{53}},
  \bibinfo{pages}{11176} (\bibinfo{year}{1996}).

\bibitem[{\citenamefont{Koslowski et~al.}(2007)\citenamefont{Koslowski, Dietrich,
  Tschetschetkin, and Ziemann}}]{koslowski07}
\bibinfo{author}{\bibfnamefont{B.} \bibnamefont{Koslowski}},
  \bibinfo{author}{\bibfnamefont{C.} \bibnamefont{Dietrich}},
  \bibinfo{author}{\bibfnamefont{A.} \bibnamefont{Tschetschetkin}}, \bibnamefont{and}
  \bibinfo{author}{\bibfnamefont{P.} \bibnamefont{Ziemann}},
  \bibinfo{journal}{Phys. Rev. B} \textbf{\bibinfo{volume}{75}},
  \bibinfo{pages}{035421} (\bibinfo{year}{2007}).

\bibitem[{\citenamefont{Passoni et~al.}(2009)\citenamefont{Passoni, Donati,
  Li Bassi, Casari, and Bottani}}]{passoni09}
\bibinfo{author}{\bibfnamefont{M.} \bibnamefont{Passoni}},
  \bibinfo{author}{\bibfnamefont{F.} \bibnamefont{Donati}},
  \bibinfo{author}{\bibfnamefont{A.} \bibnamefont{Li~Bassi}},
  \bibinfo{author}{\bibfnamefont{C.~S.} \bibnamefont{Casari}}, \bibnamefont{and}
  \bibinfo{author}{\bibfnamefont{C.~E.} \bibnamefont{Bottani}},
  \bibinfo{journal}{Phys. Rev. B} \textbf{\bibinfo{volume}{79}},
  \bibinfo{pages}{045404} (\bibinfo{year}{2009}).

\bibitem[{\citenamefont{Ziegler et~al.}(2009)\citenamefont{Ziegler, N\'eel,
  Sperl, Kr\"oger, and Berndt}}]{ziegler09}
\bibinfo{author}{\bibfnamefont{M.} \bibnamefont{Ziegler}},
  \bibinfo{author}{\bibfnamefont{N.} \bibnamefont{N\'eel}},
  \bibinfo{author}{\bibfnamefont{A.} \bibnamefont{Sperl}},
  \bibinfo{author}{\bibfnamefont{J.} \bibnamefont{Kr\"oger}}, \bibnamefont{and}
  \bibinfo{author}{\bibfnamefont{R.} \bibnamefont{Berndt}},
  \bibinfo{journal}{Phys. Rev. B} \textbf{\bibinfo{volume}{80}},
  \bibinfo{pages}{125402} (\bibinfo{year}{2009}).

\bibitem[{\citenamefont{Koslowski et~al.}(2009)\citenamefont{Koslowski, Pfeifer,
  and Ziemann}}]{koslowski09}
\bibinfo{author}{\bibfnamefont{B.} \bibnamefont{Koslowski}},
  \bibinfo{author}{\bibfnamefont{H.} \bibnamefont{Pfeifer}}, \bibnamefont{and}
  \bibinfo{author}{\bibfnamefont{P.} \bibnamefont{Ziemann}},
  \bibinfo{journal}{Phys. Rev. B} \textbf{\bibinfo{volume}{80}},
  \bibinfo{pages}{165419} (\bibinfo{year}{2009}).

\bibitem[{\citenamefont{Kwapi\'nski and Ja\l ochowski}(2010)}]{kwapinski10}
\bibinfo{author}{\bibfnamefont{T.}~\bibnamefont{Kwapi\'nski}} \bibnamefont{and}
  \bibinfo{author}{\bibfnamefont{M.}~\bibnamefont{Ja\l ochowski}},
  \bibinfo{journal}{Surf. Sci.} \textbf{\bibinfo{volume}{604}},
  \bibinfo{pages}{1752} (\bibinfo{year}{2010}).

\bibitem[{\citenamefont{Hofer and Garcia-Lekue}(2005)}]{hofer05sts}
\bibinfo{author}{\bibfnamefont{W.~A.}~\bibnamefont{Hofer}} \bibnamefont{and}
  \bibinfo{author}{\bibfnamefont{A.}~\bibnamefont{Garcia-Lekue}},
  \bibinfo{journal}{Phys. Rev. B} \textbf{\bibinfo{volume}{71}},
  \bibinfo{pages}{085401} (\bibinfo{year}{2005}).

\bibitem[{\citenamefont{Plumer et~al.}(2001)\citenamefont{Plumer, van Ek, and
  Weller}}]{ultrahigh}
\bibinfo{author}{\bibfnamefont{E.~M.~L.} \bibnamefont{Plumer}},
  \bibinfo{author}{\bibfnamefont{J.}~\bibnamefont{van Ek}}, \bibnamefont{and}
  \bibinfo{author}{\bibfnamefont{D.}~\bibnamefont{Weller}},
  \emph{\bibinfo{title}{The Physics of Ultra-High Density Magnetic Recording,
  Springer Series in Surface Science Vol. 41}} (\bibinfo{publisher}{Springer},
  \bibinfo{address}{Berlin, Germany}, \bibinfo{year}{2001}).

\bibitem[{\citenamefont{Weiss et~al.}(2005)\citenamefont{Weiss, Cren, Epple,
  Rusponi, Baudot, Rohart, Tejeda, Repain, Rousset, Ohresser et~al.}}]{weiss05}
\bibinfo{author}{\bibfnamefont{N.}~\bibnamefont{Weiss}},
  \bibinfo{author}{\bibfnamefont{T.}~\bibnamefont{Cren}},
  \bibinfo{author}{\bibfnamefont{M.}~\bibnamefont{Epple}},
  \bibinfo{author}{\bibfnamefont{S.}~\bibnamefont{Rusponi}},
  \bibinfo{author}{\bibfnamefont{G.}~\bibnamefont{Baudot}},
  \bibinfo{author}{\bibfnamefont{S.}~\bibnamefont{Rohart}},
  \bibinfo{author}{\bibfnamefont{A.}~\bibnamefont{Tejeda}},
  \bibinfo{author}{\bibfnamefont{V.}~\bibnamefont{Repain}},
  \bibinfo{author}{\bibfnamefont{S.}~\bibnamefont{Rousset}},
  \bibinfo{author}{\bibfnamefont{P.}~\bibnamefont{Ohresser}},
  \bibinfo{author}{\bibfnamefont{F.}~\bibnamefont{Scheurer}},
  \bibinfo{author}{\bibfnamefont{P.}~\bibnamefont{Bencok}}, \bibnamefont{and}
  \bibinfo{author}{\bibfnamefont{H.}~\bibnamefont{Brune}},
  \bibinfo{journal}{Phys. Rev. Lett.} \textbf{\bibinfo{volume}{95}},
  \bibinfo{pages}{157204} (\bibinfo{year}{2005}).

\bibitem[{\citenamefont{Bode}(2003)}]{bode03review}
\bibinfo{author}{\bibfnamefont{M.}~\bibnamefont{Bode}},
  \bibinfo{journal}{Rep. Prog. Phys.} \textbf{\bibinfo{volume}{66}},
  \bibinfo{pages}{523} (\bibinfo{year}{2003}).

\bibitem[{\citenamefont{Wiesendanger}(2009)}]{wiesendanger09review}
\bibinfo{author}{\bibfnamefont{R.}~\bibnamefont{Wiesendanger}},
  \bibinfo{journal}{Rev. Mod. Phys.} \textbf{\bibinfo{volume}{81}},
  \bibinfo{pages}{1495} (\bibinfo{year}{2009}).

\bibitem[{\citenamefont{Wulfhekel and Gao}(2010)}]{wulfhekel10review}
\bibinfo{author}{\bibfnamefont{W.}~\bibnamefont{Wulfhekel}} \bibnamefont{and}
  \bibinfo{author}{\bibfnamefont{C.~L.}~\bibnamefont{Gao}},
  \bibinfo{journal}{J. Phys. Condens. Matter} \textbf{\bibinfo{volume}{22}},
  \bibinfo{pages}{084021} (\bibinfo{year}{2010}).

\bibitem[{\citenamefont{Serrate et~al.}(2010)\citenamefont{Serrate, Ferriani,
  Yoshida, Hla, Menzel, Bergmann, Heinze, Kubetzka, and Wiesendanger}}]{serrate10}
\bibinfo{author}{\bibfnamefont{D.}~\bibnamefont{Serrate}},
  \bibinfo{author}{\bibfnamefont{P.}~\bibnamefont{Ferriani}},
  \bibinfo{author}{\bibfnamefont{Y.}~\bibnamefont{Yoshida}},
  \bibinfo{author}{\bibfnamefont{S.-W.}~\bibnamefont{Hla}},
  \bibinfo{author}{\bibfnamefont{M.}~\bibnamefont{Menzel}},
  \bibinfo{author}{\bibfnamefont{K.~von}~\bibnamefont{Bergmann}},
  \bibinfo{author}{\bibfnamefont{S.}~\bibnamefont{Heinze}},
  \bibinfo{author}{\bibfnamefont{A.}~\bibnamefont{Kubetzka}}, \bibnamefont{and}
  \bibinfo{author}{\bibfnamefont{R.}~\bibnamefont{Wiesendanger}},
  \bibinfo{journal}{Nature Nanotechnology} \textbf{\bibinfo{volume}{5}},
  \bibinfo{pages}{350} (\bibinfo{year}{2010}).

\bibitem[{\citenamefont{Heinze et~al.}(2011)\citenamefont{Heinze, von Bergmann,
  Menzel, Brede, Kubetzka, Wiesendanger, Bihlmayer, and Bl\"ugel}}]{heinze11skyrmion}
\bibinfo{author}{\bibfnamefont{S.}~\bibnamefont{Heinze}},
  \bibinfo{author}{\bibfnamefont{K.}~\bibnamefont{von}~\bibnamefont{Bergmann}},
  \bibinfo{author}{\bibfnamefont{M.}~\bibnamefont{Menzel}},
  \bibinfo{author}{\bibfnamefont{J.}~\bibnamefont{Brede}},
  \bibinfo{author}{\bibfnamefont{A.}~\bibnamefont{Kubetzka}},
  \bibinfo{author}{\bibfnamefont{R.}~\bibnamefont{Wiesendanger}},
  \bibinfo{author}{\bibfnamefont{G.}~\bibnamefont{Bihlmayer}}, \bibnamefont{and}
  \bibinfo{author}{\bibfnamefont{S.}~\bibnamefont{Bl\"ugel}},
  \bibinfo{journal}{Nature Physics} \textbf{\bibinfo{volume}{7}},
  \bibinfo{pages}{713} (\bibinfo{year}{2011}).

\bibitem[{\citenamefont{Yayon et~al.}(2007)\citenamefont{Yayon, Brar,
  Senapati, Erwin, and Crommie}}]{yayon07}
\bibinfo{author}{\bibfnamefont{Y.}~\bibnamefont{Yayon}},
  \bibinfo{author}{\bibfnamefont{V.~W.}~\bibnamefont{Brar}},
  \bibinfo{author}{\bibfnamefont{L.}~\bibnamefont{Senapati}},
  \bibinfo{author}{\bibfnamefont{S.~C.}~\bibnamefont{Erwin}}, \bibnamefont{and}
  \bibinfo{author}{\bibfnamefont{M.~F.}~\bibnamefont{Crommie}},
  \bibinfo{journal}{Phys. Rev. Lett.} \textbf{\bibinfo{volume}{99}},
  \bibinfo{pages}{067202} (\bibinfo{year}{2007}).

\bibitem[{\citenamefont{Heinrich et~al.}(2009)\citenamefont{Heinrich, Iacovita,
  Rastei, Limot, Bucher, Ignatiev, Stepanyuk, and Bruno}}]{heinrich09}
\bibinfo{author}{\bibfnamefont{B.~W.}~\bibnamefont{Heinrich}},
  \bibinfo{author}{\bibfnamefont{C.}~\bibnamefont{Iacovita}},
  \bibinfo{author}{\bibfnamefont{M.~V.}~\bibnamefont{Rastei}},
  \bibinfo{author}{\bibfnamefont{L.}~\bibnamefont{Limot}},
  \bibinfo{author}{\bibfnamefont{J.~P.}~\bibnamefont{Bucher}},
  \bibinfo{author}{\bibfnamefont{P.~A.}~\bibnamefont{Ignatiev}},
  \bibinfo{author}{\bibfnamefont{V.~S.}~\bibnamefont{Stepanyuk}}, \bibnamefont{and}
  \bibinfo{author}{\bibfnamefont{P.}~\bibnamefont{Bruno}},
  \bibinfo{journal}{Phys. Rev. B} \textbf{\bibinfo{volume}{79}},
  \bibinfo{pages}{113401} (\bibinfo{year}{2009}).

\bibitem[{\citenamefont{Zhou et~al.}(2010)\citenamefont{Zhou, Meier,
  Wiebe, and Wiesendanger}}]{zhou10}
\bibinfo{author}{\bibfnamefont{L.} \bibnamefont{Zhou}},
  \bibinfo{author}{\bibfnamefont{F.} \bibnamefont{Meier}},
  \bibinfo{author}{\bibfnamefont{J.} \bibnamefont{Wiebe}}, \bibnamefont{and}
  \bibinfo{author}{\bibfnamefont{R.} \bibnamefont{Wiesendanger}},
  \bibinfo{journal}{Phys. Rev. B} \textbf{\bibinfo{volume}{82}},
  \bibinfo{pages}{012409} (\bibinfo{year}{2010}).

\bibitem[{\citenamefont{Ferriani et~al.}(2010)\citenamefont{Ferriani, Lazo,
  and Heinze}}]{ferriani10tip}
\bibinfo{author}{\bibfnamefont{P.} \bibnamefont{Ferriani}},
  \bibinfo{author}{\bibfnamefont{C.} \bibnamefont{Lazo}}, \bibnamefont{and}
  \bibinfo{author}{\bibfnamefont{S.} \bibnamefont{Heinze}},
  \bibinfo{journal}{Phys. Rev. B} \textbf{\bibinfo{volume}{82}},
  \bibinfo{pages}{054411} (\bibinfo{year}{2010}).

\bibitem[{\citenamefont{Wiebe et~al.}(2011)\citenamefont{Wiebe, Zhou,
  and Wiesendanger}}]{wiebe11}
\bibinfo{author}{\bibfnamefont{J.} \bibnamefont{Wiebe}},
  \bibinfo{author}{\bibfnamefont{L.} \bibnamefont{Zhou}}, \bibnamefont{and}
  \bibinfo{author}{\bibfnamefont{R.} \bibnamefont{Wiesendanger}},
  \bibinfo{journal}{J. Phys. D: Appl. Phys.} \textbf{\bibinfo{volume}{44}},
  \bibinfo{pages}{464009} (\bibinfo{year}{2011}).

\bibitem[{\citenamefont{N\'eel et~al.}(2010)\citenamefont{N\'eel, Kr\"oger,
  and Berndt}}]{neel10kondo}
\bibinfo{author}{\bibfnamefont{N.}~\bibnamefont{N\'eel}},
  \bibinfo{author}{\bibfnamefont{J.}~\bibnamefont{Kr\"oger}}, \bibnamefont{and}
  \bibinfo{author}{\bibfnamefont{R.}~\bibnamefont{Berndt}},
  \bibinfo{journal}{Phys. Rev. B} \textbf{\bibinfo{volume}{82}},
  \bibinfo{pages}{233401} (\bibinfo{year}{2010}).

\bibitem[{\citenamefont{Ternes et~al.}(2009)\citenamefont{Ternes, Heinrich,
  and Schneider}}]{ternes09sts}
\bibinfo{author}{\bibfnamefont{M.}~\bibnamefont{Ternes}},
  \bibinfo{author}{\bibfnamefont{A.~J.}~\bibnamefont{Heinrich}}, \bibnamefont{and}
  \bibinfo{author}{\bibfnamefont{W.-D.}~\bibnamefont{Schneider}},
  \bibinfo{journal}{J. Phys. Condens. Matter} \textbf{\bibinfo{volume}{21}},
  \bibinfo{pages}{053001} (\bibinfo{year}{2009}).

\bibitem[{\citenamefont{Schouteden et~al.}(2008)\citenamefont{Schouteden,
  Muzychenko, and Van Haesendonck}}]{schouteden08}
\bibinfo{author}{\bibfnamefont{K.}~\bibnamefont{Schouteden}},
  \bibinfo{author}{\bibfnamefont{D.~A.}~\bibnamefont{Muzychenko}}, \bibnamefont{and}
  \bibinfo{author}{\bibfnamefont{C.}~\bibnamefont{Van~Haesendonck}},
  \bibinfo{journal}{J. Nanosci. Nanotechnol.} \textbf{\bibinfo{volume}{8}},
  \bibinfo{pages}{3616} (\bibinfo{year}{2008}).

\bibitem[{\citenamefont{Heinrich et~al.}(2010)\citenamefont{Heinrich, Iacovita,
  Rastei, Limot, Ignatiev, Stepanyuk, and Bucher}}]{heinrich10}
\bibinfo{author}{\bibfnamefont{B.~W.}~\bibnamefont{Heinrich}},
  \bibinfo{author}{\bibfnamefont{C.}~\bibnamefont{Iacovita}},
  \bibinfo{author}{\bibfnamefont{M.~V.}~\bibnamefont{Rastei}},
  \bibinfo{author}{\bibfnamefont{L.}~\bibnamefont{Limot}},
  \bibinfo{author}{\bibfnamefont{P.~A.}~\bibnamefont{Ignatiev}},
  \bibinfo{author}{\bibfnamefont{V.~S.}~\bibnamefont{Stepanyuk}}, \bibnamefont{and}
  \bibinfo{author}{\bibfnamefont{J.~P.}~\bibnamefont{Bucher}},
  \bibinfo{journal}{Eur. Phys. J. B} \textbf{\bibinfo{volume}{75}},
  \bibinfo{pages}{49} (\bibinfo{year}{2010}).

\bibitem[{\citenamefont{Rodary et~al.}(2009)\citenamefont{Rodary, Wedekind,
  Oka, Sander, and Kirschner}}]{rodary09}
\bibinfo{author}{\bibfnamefont{G.}~\bibnamefont{Rodary}},
  \bibinfo{author}{\bibfnamefont{S.}~\bibnamefont{Wedekind}},
  \bibinfo{author}{\bibfnamefont{H.}~\bibnamefont{Oka}},
  \bibinfo{author}{\bibfnamefont{D.}~\bibnamefont{Sander}}, \bibnamefont{and}
  \bibinfo{author}{\bibfnamefont{J.}~\bibnamefont{Kirschner}},
  \bibinfo{journal}{Appl. Phys. Lett.} \textbf{\bibinfo{volume}{95}},
  \bibinfo{pages}{152513} (\bibinfo{year}{2009}).

\bibitem[{\citenamefont{Palot\'as et~al.}(2011)\citenamefont{Palot\'as, Hofer,
  and Szunyogh}}]{palotas11sts}
\bibinfo{author}{\bibfnamefont{K.}~\bibnamefont{Palot\'as}},
  \bibinfo{author}{\bibfnamefont{W.~A.}~\bibnamefont{Hofer}}, \bibnamefont{and}
  \bibinfo{author}{\bibfnamefont{L.}~\bibnamefont{Szunyogh}},
  \bibinfo{journal}{Phys. Rev. B} \textbf{\bibinfo{volume}{83}},
  \bibinfo{pages}{214410} (\bibinfo{year}{2011}).

\bibitem[{\citenamefont{Hofer et~al.}(2008)\citenamefont{Hofer, Palot\'as,
  Rusponi, Cren, and Brune}}]{hofer08tipH}
\bibinfo{author}{\bibfnamefont{W.~A.}~\bibnamefont{Hofer}},
  \bibinfo{author}{\bibfnamefont{K.}~\bibnamefont{Palot\'as}},
  \bibinfo{author}{\bibfnamefont{S.}~\bibnamefont{Rusponi}},
  \bibinfo{author}{\bibfnamefont{T.}~\bibnamefont{Cren}}, \bibnamefont{and}
  \bibinfo{author}{\bibfnamefont{H.}~\bibnamefont{Brune}},
  \bibinfo{journal}{Phys. Rev. Lett.} \textbf{\bibinfo{volume}{100}},
  \bibinfo{pages}{026806} (\bibinfo{year}{2008}).

\bibitem[{\citenamefont{Palot\'as et~al.}(2011)\citenamefont{Palot\'as, Hofer,
  and Szunyogh}}]{palotas11stm}
\bibinfo{author}{\bibfnamefont{K.}~\bibnamefont{Palot\'as}},
  \bibinfo{author}{\bibfnamefont{W.~A.}~\bibnamefont{Hofer}}, \bibnamefont{and}
  \bibinfo{author}{\bibfnamefont{L.}~\bibnamefont{Szunyogh}},
  \bibinfo{journal}{Phys. Rev. B} \textbf{\bibinfo{volume}{84}},
  \bibinfo{pages}{174428} (\bibinfo{year}{2011}).

\bibitem[{\citenamefont{Wortmann et~al.}(2001)\citenamefont{Wortmann,
  Heinze, Kurz, Bihlmayer, and Bl\"ugel}}]{wortmann01}
\bibinfo{author}{\bibfnamefont{D.}~\bibnamefont{Wortmann}},
  \bibinfo{author}{\bibfnamefont{S.}~\bibnamefont{Heinze}},
  \bibinfo{author}{\bibfnamefont{P.}~\bibnamefont{Kurz}},
  \bibinfo{author}{\bibfnamefont{G.}~\bibnamefont{Bihlmayer}}, \bibnamefont{and}
  \bibinfo{author}{\bibfnamefont{S.}~\bibnamefont{Bl\"ugel}},
  \bibinfo{journal}{Phys. Rev. Lett.} \textbf{\bibinfo{volume}{86}},
  \bibinfo{pages}{4132} (\bibinfo{year}{2001}).

\bibitem[{\citenamefont{Heinze}(2006)}]{heinze06}
\bibinfo{author}{\bibfnamefont{S.}~\bibnamefont{Heinze}},
  \bibinfo{journal}{Appl. Phys. A} \textbf{\bibinfo{volume}{85}},
  \bibinfo{pages}{407} (\bibinfo{year}{2006}).

\bibitem[{\citenamefont{Donati et~al.}(2011)\citenamefont{Donati, Piccoli,
  Bottani, and Passoni}}]{donati11}
\bibinfo{author}{\bibfnamefont{F.}~\bibnamefont{Donati}},
  \bibinfo{author}{\bibfnamefont{S.}~\bibnamefont{Piccoli}},
  \bibinfo{author}{\bibfnamefont{C.~E.}~\bibnamefont{Bottani}}, \bibnamefont{and}
  \bibinfo{author}{\bibfnamefont{M.}~\bibnamefont{Passoni}},
  \bibinfo{journal}{New J. Phys.} \textbf{\bibinfo{volume}{13}},
  \bibinfo{pages}{053058} (\bibinfo{year}{2011}).

\bibitem[{\citenamefont{Passoni and Bottani}(2007)}]{passoni07}
\bibinfo{author}{\bibfnamefont{M.}~\bibnamefont{Passoni}} \bibnamefont{and}
  \bibinfo{author}{\bibfnamefont{C.~E.}~\bibnamefont{Bottani}},
  \bibinfo{journal}{Phys. Rev. B} \textbf{\bibinfo{volume}{76}},
  \bibinfo{pages}{115404} (\bibinfo{year}{2007}).

\bibitem[{\citenamefont{Yang et~al.}(2002)\citenamefont{Yang,
  Smith, Prikhodko, and Lambrecht}}]{yang02}
\bibinfo{author}{\bibfnamefont{H.}~\bibnamefont{Yang}},
  \bibinfo{author}{\bibfnamefont{A.~R.}~\bibnamefont{Smith}},
  \bibinfo{author}{\bibfnamefont{M.}~\bibnamefont{Prikhodko}}, \bibnamefont{and}
  \bibinfo{author}{\bibfnamefont{W.~R.~L.}~\bibnamefont{Lambrecht}},
  \bibinfo{journal}{Phys. Rev. Lett.} \textbf{\bibinfo{volume}{89}},
  \bibinfo{pages}{226101} (\bibinfo{year}{2002}).

\bibitem[{\citenamefont{Smith et~al.}(2004)\citenamefont{Smith,
  Yang, Yang, Lambrecht, Dick, and Neugebauer}}]{smith04}
\bibinfo{author}{\bibfnamefont{A.~R.}~\bibnamefont{Smith}},
  \bibinfo{author}{\bibfnamefont{R.}~\bibnamefont{Yang}},
  \bibinfo{author}{\bibfnamefont{H.}~\bibnamefont{Yang}},
  \bibinfo{author}{\bibfnamefont{W.~R.~L.}~\bibnamefont{Lambrecht}},
  \bibinfo{author}{\bibfnamefont{A.}~\bibnamefont{Dick}}, \bibnamefont{and}
  \bibinfo{author}{\bibfnamefont{J.}~\bibnamefont{Neugebauer}},
  \bibinfo{journal}{Surf. Sci.} \textbf{\bibinfo{volume}{561}},
  \bibinfo{pages}{154} (\bibinfo{year}{2004}).

\bibitem[{\citenamefont{Tersoff and Hamann}(1983)}]{tersoff83}
\bibinfo{author}{\bibfnamefont{J.}~\bibnamefont{Tersoff}} \bibnamefont{and}
  \bibinfo{author}{\bibfnamefont{D.~R.}~\bibnamefont{Hamann}},
  \bibinfo{journal}{Phys. Rev. Lett.} \textbf{\bibinfo{volume}{50}},
  \bibinfo{pages}{1998} (\bibinfo{year}{1983}).

\bibitem[{\citenamefont{Tersoff and Hamann}(1985)}]{tersoff85}
\bibinfo{author}{\bibfnamefont{J.}~\bibnamefont{Tersoff}} \bibnamefont{and}
  \bibinfo{author}{\bibfnamefont{D.~R.}~\bibnamefont{Hamann}},
  \bibinfo{journal}{Phys. Rev. B} \textbf{\bibinfo{volume}{31}},
  \bibinfo{pages}{805} (\bibinfo{year}{1985}).

\bibitem[{\citenamefont{Chen}(1990)}]{chen90}
\bibinfo{author}{\bibfnamefont{C.~J.}~\bibnamefont{Chen}},
  \bibinfo{journal}{Phys. Rev. B} \textbf{\bibinfo{volume}{42}},
  \bibinfo{pages}{8841} (\bibinfo{year}{1990}).

\bibitem[{\citenamefont{Ding et~al.}(2003)\citenamefont{Ding,
  Wulfhekel, Henk, Bruno, and Kirschner}}]{ding03}
\bibinfo{author}{\bibfnamefont{H.~F.}~\bibnamefont{Ding}},
  \bibinfo{author}{\bibfnamefont{W.}~\bibnamefont{Wulfhekel}},
  \bibinfo{author}{\bibfnamefont{J.}~\bibnamefont{Henk}},
  \bibinfo{author}{\bibfnamefont{P.}~\bibnamefont{Bruno}}, \bibnamefont{and}
  \bibinfo{author}{\bibfnamefont{J.}~\bibnamefont{Kirschner}},
  \bibinfo{journal}{Phys. Rev. Lett.} \textbf{\bibinfo{volume}{90}},
  \bibinfo{pages}{116603} (\bibinfo{year}{2003}).

\bibitem[{\citenamefont{Tange et~al.}(2010)\citenamefont{Tange, Gao,
  Yavorsky, Maznichenko, Etz, Ernst, Hergert, Mertig, Wulfhekel, and Kirschner}}]{tange10}
\bibinfo{author}{\bibfnamefont{A.}~\bibnamefont{Tange}},
  \bibinfo{author}{\bibfnamefont{C.~L.}~\bibnamefont{Gao}},
  \bibinfo{author}{\bibfnamefont{B.~Y.}~\bibnamefont{Yavorsky}},
  \bibinfo{author}{\bibfnamefont{I.~V.}~\bibnamefont{Maznichenko}},
  \bibinfo{author}{\bibfnamefont{C.}~\bibnamefont{Etz}},
  \bibinfo{author}{\bibfnamefont{A.}~\bibnamefont{Ernst}},
  \bibinfo{author}{\bibfnamefont{W.}~\bibnamefont{Hergert}},
  \bibinfo{author}{\bibfnamefont{I.}~\bibnamefont{Mertig}},
  \bibinfo{author}{\bibfnamefont{W.}~\bibnamefont{Wulfhekel}}, \bibnamefont{and}
  \bibinfo{author}{\bibfnamefont{J.}~\bibnamefont{Kirschner}},
  \bibinfo{journal}{Phys. Rev. B} \textbf{\bibinfo{volume}{81}},
  \bibinfo{pages}{195410} (\bibinfo{year}{2010}).

\bibitem[{\citenamefont{Kresse and Furthm\"uller}(1996{\natexlab{a}})}]{VASP2}
\bibinfo{author}{\bibfnamefont{G.}~\bibnamefont{Kresse}} \bibnamefont{and}
  \bibinfo{author}{\bibfnamefont{J.}~\bibnamefont{Furthm\"uller}},
  \bibinfo{journal}{Comput. Mater. Sci.} \textbf{\bibinfo{volume}{6}},
  \bibinfo{pages}{15} (\bibinfo{year}{1996}{\natexlab{a}}).

\bibitem[{\citenamefont{Kresse and Furthm\"uller}(1996{\natexlab{b}})}]{VASP3}
\bibinfo{author}{\bibfnamefont{G.}~\bibnamefont{Kresse}} \bibnamefont{and}
  \bibinfo{author}{\bibfnamefont{J.}~\bibnamefont{Furthm\"uller}},
  \bibinfo{journal}{Phys. Rev. B} \textbf{\bibinfo{volume}{54}},
  \bibinfo{pages}{11169} (\bibinfo{year}{1996}{\natexlab{b}}).

\bibitem[{\citenamefont{Hafner}(2008)}]{hafner08}
\bibinfo{author}{\bibfnamefont{J.}~\bibnamefont{Hafner}},
  \bibinfo{journal}{J. Comput. Chem.} \textbf{\bibinfo{volume}{29}},
  \bibinfo{pages}{2044} (\bibinfo{year}{2008}).

\bibitem[{\citenamefont{Kresse and Joubert}(1999)}]{kresse99}
\bibinfo{author}{\bibfnamefont{G.}~\bibnamefont{Kresse}} \bibnamefont{and}
  \bibinfo{author}{\bibfnamefont{D.}~\bibnamefont{Joubert}},
  \bibinfo{journal}{Phys. Rev. B} \textbf{\bibinfo{volume}{59}},
  \bibinfo{pages}{1758} (\bibinfo{year}{1999}).

\bibitem[{\citenamefont{Perdew and Wang}(1992)}]{pw91}
\bibinfo{author}{\bibfnamefont{J.~P.} \bibnamefont{Perdew}} \bibnamefont{and}
  \bibinfo{author}{\bibfnamefont{Y.}~\bibnamefont{Wang}},
  \bibinfo{journal}{Phys. Rev. B} \textbf{\bibinfo{volume}{45}},
  \bibinfo{pages}{13244} (\bibinfo{year}{1992}).

\bibitem[{\citenamefont{Hobbs et~al.}(2000)\citenamefont{Hobbs, Kresse,
  and Hafner}}]{hobbs00prb}
\bibinfo{author}{\bibfnamefont{D.}~\bibnamefont{Hobbs}},
  \bibinfo{author}{\bibfnamefont{G.}~\bibnamefont{Kresse}}, \bibnamefont{and}
  \bibinfo{author}{\bibfnamefont{J.}~\bibnamefont{Hafner}},
  \bibinfo{journal}{Phys. Rev. B} \textbf{\bibinfo{volume}{62}},
  \bibinfo{pages}{11556} (\bibinfo{year}{2000}).

\bibitem[{\citenamefont{Hobbs and Hafner}(2000)}]{hobbs00jpcm}
\bibinfo{author}{\bibfnamefont{D.}~\bibnamefont{Hobbs}} \bibnamefont{and}
  \bibinfo{author}{\bibfnamefont{J.}~\bibnamefont{Hafner}},
  \bibinfo{journal}{J. Phys. Condens. Matter} \textbf{\bibinfo{volume}{12}},
  \bibinfo{pages}{7025} (\bibinfo{year}{2000}).

\bibitem[{\citenamefont{Monkhorst and Pack}(1976)}]{monkhorst}
\bibinfo{author}{\bibfnamefont{H.~J.} \bibnamefont{Monkhorst}} \bibnamefont{and}
  \bibinfo{author}{\bibfnamefont{J.~D.}~\bibnamefont{Pack}},
  \bibinfo{journal}{Phys. Rev. B} \textbf{\bibinfo{volume}{13}},
  \bibinfo{pages}{5188} (\bibinfo{year}{1976}).

\bibitem[{\citenamefont{Kubetzka et~al.}(2005)\citenamefont{Kubetzka, Ferriani,
  Bode, Heinze, Bihlmayer, Bergmann, Pietzsch, Bl\"ugel, and Wiesendanger}}]{kubetzka05}
\bibinfo{author}{\bibfnamefont{A.}~\bibnamefont{Kubetzka}},
  \bibinfo{author}{\bibfnamefont{P.}~\bibnamefont{Ferriani}},
  \bibinfo{author}{\bibfnamefont{M.}~\bibnamefont{Bode}},
  \bibinfo{author}{\bibfnamefont{S.}~\bibnamefont{Heinze}},
  \bibinfo{author}{\bibfnamefont{G.}~\bibnamefont{Bihlmayer}},
  \bibinfo{author}{\bibfnamefont{K.~von}~\bibnamefont{Bergmann}},
  \bibinfo{author}{\bibfnamefont{O.}~\bibnamefont{Pietzsch}},
  \bibinfo{author}{\bibfnamefont{S.}~\bibnamefont{Bl\"ugel}}, \bibnamefont{and}
  \bibinfo{author}{\bibfnamefont{R.}~\bibnamefont{Wiesendanger}},
  \bibinfo{journal}{Phys. Rev. Lett.} \textbf{\bibinfo{volume}{94}},
  \bibinfo{pages}{087204} (\bibinfo{year}{2005}).


\end{thebibliography}
\end{document}